\documentclass[jcp,aip,preprint,onecolumn]{revtex4-1}
\usepackage[colorlinks={true},dvipdfm]{hyperref}
\hypersetup{citecolor={blue}, filecolor={blue}, linkcolor={blue}, urlcolor={blue}}
\usepackage{anysize}
\usepackage{graphicx}
\usepackage{times}
\marginsize{1 in}{1 in}{1 in}{1 in}
\usepackage{amsmath,amssymb}
\usepackage{dcolumn}
\usepackage{bm}
\usepackage{enumerate}
\newcommand{\pif}{$P_{i\to f}$}
\newcommand{\mpif}{P_{i\to f}}
\newcommand{\mpin}{P_{1\to N}}
\newcommand{\ms}{\medspace}
\newcommand{\dg}{^{\dag}}
\newcommand{\tprime}{t^{\prime}}
\newcommand{\mprime}{m^{\prime}}

\newcommand{\e}{\varepsilon}

\begin{document}

\title{Exploring Constrained Quantum Control Landscapes}
\author{Katharine W. Moore}
\author{Herschel Rabitz}
\affiliation{Department of Chemistry, Princeton University, Princeton, NJ 08544, USA}
\date{\today}

\begin{abstract}
The broad success of optimally controlling quantum systems with external fields has been attributed to the favorable topology of the underlying control landscape, where the landscape is the physical observable as a function of the controls. The control landscape can be shown to contain no suboptimal trapping extrema upon satisfaction of reasonable physical assumptions, but this topological analysis does not hold when significant constraints are placed on the control resources. This work employs simulations to explore the topology and features of the control landscape for pure-state population transfer with a constrained class of control fields. The fields are parameterized in terms of a set of uniformly spaced spectral frequencies, with the associated phases acting as the controls. This restricted family of fields provides a simple illustration for assessing the impact of constraints upon seeking optimal control. Optimization results reveal that the minimum number of phase controls necessary to assure a high yield in the target state has a special dependence on the number of accessible energy levels in the quantum system, revealed from an analysis of the first- and second-order variation of the yield with respect to the controls. When an insufficient number of controls and/or a weak control fluence are employed, trapping extrema and saddle points are observed on the landscape. When the control resources are sufficiently flexible, solutions producing the globally maximal yield are found to form connected `level sets' of continuously variable control fields that preserve the yield. These optimal yield level sets are found to shrink to isolated points on the top of the landscape as the control field fluence is decreased, and further reduction of the fluence turns these points into suboptimal trapping extrema on the landscape. Although constrained control fields can come in many forms beyond the cases explored here, the behavior found in this paper is illustrative of the impacts that constraints can introduce.
\end{abstract}
\maketitle

\section{Introduction}
The control of quantum systems with tailored external laser fields is an active area of research. Optimal control experiments (OCE) employing closed-loop learning control \cite{judson} have found success in a wide range of applications \cite{constantin}, including high harmonic generation \cite{Bartels2000, Bartels2004, Pfeifer2005}, bond-breaking in molecules \cite{Levis2001, Gerber02,Gerber2002, Weinacht2005b,melab2}, and energy transfer in biomolecules \cite{herek}. Simulations employing optimal control theory (OCT) have achieved high yields in large numbers of quantum systems \cite{constantin,raj}. This wide success is attributed to the inherent attractive topology of the underlying {\it quantum control landscape}, which is the functional relationship between the targeted objective (e.g., transfer to an excited state, breaking a chemical bond) and the control field. Control landscapes for finite-dimensional quantum systems possess a {\it trap-free} topology \cite{raj,mike1, demiralp, taksan}, with no sub-optimal local extrema that can hinder attainment of the optimal objective value, upon satisfaction of the Assumptions that: (a) the target quantum system is controllable \cite{ramakrishna}, (b) the map from the space of control fields to the associated dynamical propagator is surjective \cite{raj,rebing}, and (c) the controls are sufficiently flexible such that the landscape may be freely traversed \cite{mike1,demiralp,taksan}.  Although uncontrollable quantum systems that violate Assumption (a) can be found, they form a null set in the space of Hamiltonians \cite{altafini}. Unusual exceptions to Assumption (b) can be found that produce landscape traps \cite{schirmer2,tannor2,rebing,comment,commentreply}, but extensive numerical simulations with a broad variety of systems that avoid constraints on the controls show that extremely high yields can be achieved without encountering traps on the landscape \cite{mepif,mew,greg}. The latter studies demonstrate the importance of Assumption (c) to operate without constraints on the control field in order to ensure a trap-free landscape topology.

Exploring the consequences of constraining the control field in a systematic fashion has only received limited attention in the context of landscape analysis \cite{ashley}, even though constraints inevitably arise in both simulations and experiments. In simulations, the control field may be freely varied, but numerical implementation will introduce some form of constraints. For example, time is typically discretized into $\sim10^2\textendash10^4$ values with the field at these time-points acting as independent controls, which reduces the continuous infinite-dimensional applied field over time to a finite-dimensional set of controls. False traps, i.e., apparent local extrema caused by constraints, have been observed on the landscape if insufficient time-resolution is employed \cite{mepif}. Another constraint commonly imposed in simulations is to limit the control field fluence in order to prevent convergence to physically unrealistic strong control fields. However, the imposition of a stringent fluence constraint can prevent attainment of very high yields \cite{raj,ren,artamonov}. In the laboratory, the fluence and bandwidth of ultrafast laser pulses are inevitably limited, and the laser bandwidth is often discretized by employing a spatial light modulator (SLM) \cite{weiner} that typically provides $\sim$100 control variables, with each taking on $\sim$100 values. Often, the dimensionality of the control space is further reduced \cite{mizoguchi,Weinacht2005b,melab2} in an attempt to accelerate the algorithmic search. For some objectives, fewer than $\sim$10 well-chosen independent controls can still achieve good yields in the laboratory \cite{Weinacht2005b,melab2}, although it may not be evident {\it a priori} which set of limited controls is satisfactory.

This work examines the consequences of choosing a constrained parameterization of the control field on the topology and features of the control landscape for pure state population transfer. We employ a control field $\e(t)$ containing a set of spectral frequencies $\omega_m$, 
\begin{equation}
\e(t)=A(t)\sum_{m} \cos(\omega_mt+\phi_m),\label{et1}
\end{equation}
where $A(t)\geq0$ is a Gaussian amplitude function. The phase $\phi_m$ associated each frequency $\omega_m$ acts as an independent control, while $A(t)$ and the set of $\omega_m$ are fixed throughout each optimization. The frequencies are chosen to be in a bandwidth determined by the transition frequencies of the field-free Hamiltonian. This choice of the field form was made to permit attainment of high control yields even when employing a small number of variables (hereafter referred to as ``controls''), while also facilitating a systematic reduction in the number of controls to explore the impact of significant constraints. The fluence of the control field in Eq. (\ref{et1}) is determined solely by $A(t)$, which is fixed throughout each optimization trajectory. This formulation is analogous to the situation often employed in the laboratory, where many OCE studies vary the spectral phase of the ultrafast laser pulse while fixing the pulse energy \cite{Gerber02,Gerber2002,Weinacht2005b,melab2}. Laser radiation with discrete spectral frequencies and a fixed temporal envelope $A(t)$, as utilized in this work, may be produced in the laboratory using a laser frequency-comb \cite{diddams}, and the relative phase associated with each frequency component of the radiation can be controlled with an SLM \cite{jiang,kirchner}. The phase controls employed here have been used in conjunction with other types of controls in simulations \cite{vinny2}, and additional physically reasonable formulations of constrained control fields could be constructed as well.

The control objective in this work is to maximize the probability \pif\ms of population transfer from an initial pure state $|i\rangle$ to a target pure state $|f\rangle$ of a closed quantum system undergoing unitary evolution. Analysis of the landscape topology for this objective without field constraints and satisfying Assumptions (a) and (b) above shows the existence of critical points, i.e., landscape extrema where the derivative $\delta \mpif/\delta\e(t)=0,\ms \forall t$, only for no population transfer $\mpif=0$ and optimal transfer $\mpif=1$ \cite{mike1,demiralp}. Further landscape analysis has demonstrated a special dependence of the first- and second-order variation of \pif\ms with respect to the controls on the dimension of the landscape Hilbert space $N$ (i.e., the number of accessible energy levels of the system). The gradient $\delta \mpif/\delta\e(t)$ may be expressed in terms of $2N-2$ linearly independent functions of time \cite{mike2}, and analysis of the Hessian matrix $\delta^2 \mpif/\delta\e(t)\delta\e(t')$ at the optimum $\mpif=1$ shows that the maximum number of eigenfunctions (i.e., with corresponding non-zero negative eigenvalues) specifying control paths that lead down off the top of the landscape is also $2N-2$ \cite{demiralp,mike2,vinny2}, which has been verified numerically \cite{shen,mepif}. These results suggest that employing at least $2N-2$ independent controls may be necessary in order to reliably attain a high yield, which was shown in a so-called {\it kinematic} formulation of the \pif\ms objective (i.e., where the controls are not dependent on the structure of the Hamiltonian) \cite{me}. This work will consider the importance of using at least $2N-2$ controls in the {\it dynamical} formulation of parameterized control fields outlined above.

The topology of both global optima and suboptimal `false traps' (i.e., local extrema arising from constraints on the controls) are of interest for designing algorithmic procedures that efficiently find optimal values of the controls over constrained landscapes. In order to fully assess the topology of the constrained landscape upon the restricted control field formulation, this work exclusively employs a local gradient-based search algorithm, which will stop climbing at a suboptimal value of \pif\ms if a trap is encountered. Global search algorithms (e.g., genetic algorithms) may step over traps, making them inappropriate for assessing topology. Global search algorithms are typically employed in laboratory investigations \cite{Gerber02,Gerber2002,Weinacht2005b,melab2,Bartels2000, Bartels2004, Pfeifer2005,Levis2001,herek}, and they may continue to be favorable once a full understanding is available for the prevalence of constraint-induced trapping at suboptimal objective values. The present work assesses the prevalence of false traps on the landscape, which may facilitate the design of local and/or global search methods that can best achieve an optimal yield.

As the landscape gradient is zero at a critical point, assessment of the second-order variation of \pif\ms with respect to the controls is necessary in order to determine the topology around a critical point. Suboptimal traps and saddle points can be distinguished based on an analysis of the eigenvalues of the Hessian matrix $\delta^2 \mpif/\delta\e(t)\delta\e(t')$ at a critical point. A strictly negative semi-definite eigenvalue spectrum indicates a trap, while the presence of both positive and negative eigenvalues indicates a saddle point. The presence of traps can hinder or prevent convergence to the global optimum $\mpif=1$ with a local search algorithm. It is possible to escape from saddle points, although these features are known to slow down convergence \cite{mew, greg}. Behavior at the top of the landscape is of prime interest, where the presence of at most $2N-2$ nonzero Hessian eigenvalues produces connected optimal submanifolds, or level sets, when more than $2N-2$ suitable controls are employed and $\mpif=1$ is reachable. Level sets at the global optimum of control landscapes have been explored using a second-order algorithm that moves within the Hessian nullspace (i.e., directions specified by eigenvectors corresponding to null Hessian eigenvalues) in order to identify families of connected control fields that maintain $\mpif=1$ \cite{vinny2}. In this work, we will assess the landscape topology of both optimal and sub-optimal critical regions. Where optimal level sets at $\mpif=1$ are identified with constrained controls, the second-order procedure for traversing these critical submanifolds on the control landscape \cite{vinny2} will be employed in order to explore their features.

The remainder of the paper is organized as follows: Section \ref{meth} provides theoretical background and presents the numerical methods utilized in the simulations. Section \ref{num} explores the consequences of choosing different numbers of controls upon the probability of encountering false traps on the landscape. Section \ref{exp} examines the topology of sub-optimal critical regions on the landscape, while Section \ref{optlset} considers the features of optimal landscape regions where $\mpif=1$ even with constraints present. Finally, Section \ref{con} presents concluding remarks.

\section{Methods}\label{meth}
\subsection{Formulation of the Control Objective}
Consider a closed quantum system of $N$ eigenstates $|1\rangle,\ldots,|N\rangle$ of the field-free Hamiltonian $H_0$ with dynamics specified by the time-dependent Hamiltonian $H(t)=H_0-\mu\e(t)$, where $\mu$ is the dipole operator and $\e(t)$ is the control field. The time-evolution of the quantum state $|\psi(t)\rangle$ is given as $|\psi(t)\rangle=U(t,0)|\psi(0)\rangle$, where $U(t,0)$ is the unitary propagator evolved from time $t=0$ to time $t$, and $|\psi(0)\rangle$ is the state of the quantum system at $t=0$. The dynamics of $U(t,0)$ are governed by the time-dependent Schr\"{o}dinger equation (in dimensionless units of $\hbar\equiv 1$)
\begin{equation}
i\frac{\partial U(t,0)}{\partial t}=H(t)U(t,0),\qquad U(0,0)\equiv \mathbb{I}.\label{sgl}
\end{equation} 

The control objective is to maximize the transition probability \pif\ms of population transfer from an initial state $|i\rangle$ to a target state $|f\rangle$ of the system at time $T$,
\begin{equation}
P_{i\to f}\equiv |\langle f|U(T,0)|i\rangle|^2.\label{J}
\end{equation}
Eq. (\ref{J}) defines the control landscape for population transfer as a functional of $\e(t)$ through the dynamics induced by the Schr\"odinger equation (\ref{sgl}). We assume that the system is controllable, that is, any arbitrary unitary matrix $U(T,0)$ can be generated by a suitably chosen field $\varepsilon(t)$ at a sufficiently long final time $T$. This condition is equivalent to the requirement that the Lie algebra generated from $H_0$ and $\mu$ forms a complete set of operators \cite{ramakrishna} and $T$ is long enough to avoid hindering the dynamics. Controllability is likely satisfied for an arbitrary $N$-state quantum system, as uncontrollable quantum systems have been shown to constitute a null set in the space of Hamiltonians \cite{altafini}. The surjectivity requirement for the map between the control space and the associated dynamical propagators (i.e., the Jacobian $\delta U(T,0)/\delta \e(t)$ is full-rank) also appears to be generally satisfied for nearly all control fields \cite{raj,rebing,comment}. Upon satisfaction of these requirements, the landscape may be analytically shown to contain no suboptimal extrema \cite{mike1,demiralp}, provided that no limitations are placed on the control field. Mild constraints on the control field may still allow for reaching a fully maximal yield $\mpif=1.0$ \cite{mepif,ashley,vinny2}. However, significant constraints on the control field may introduce apparent suboptimal extrema, or false traps, on the control landscape. The encroachment of control field constraints on the nominal trap-free landscape is the topic explored in this work.

\subsection{Construction of the Hamiltonian and Control Field}\label{param}
Many different structures of the field-free Hamiltonian $H_0$ and dipole matrix $\mu$ may be employed for optimal control simulations. Several distinct control systems are considered in Refs. \cite{mepif,mew}, with the landscape topology and structure found to be qualitatively the same regardless of the Hamiltonian. For simplicity, we consider only one $H_0$ and $\mu$ structure here. The free Hamiltonian $H_0$ has energy levels like that of a rigid rotor,
\begin{equation}
H_0=\sum_{j=0}^{N-1} \thinspace j\left(j+1\right)|j\rangle\langle j|.\label{ho}
\end{equation}
In many physical systems, the dipole coupling strength between states decreases as the difference between the quantum numbers of the states increases. The dipole $\mu$ takes this property into account and has the structure
\begin{equation}
\mu=\sum_{j\neq k=0}^{N-1}\frac{1}{2^{|j-k|-4}}|j\rangle\langle k|.\label{mu}
\end{equation}

The control fields $\e(t)$ consist of $M+1$ evenly-spaced spectral frequencies $\{\omega_m\}$ of equal amplitude, where the spectral phases $\{\phi_m\}$ of the first $M$ frequencies constitute $M$ independent controls for optimization,
\begin{equation}
\e(t)=A(t)\sum_{m=1}^{M+1} \cos(\omega_mt+\phi_m).\label{et}
\end{equation}
As the physical meaning of the phase of a frequency component in a spectrum is only defined with respect to the phase of a reference spectral component, we choose $\phi_{M+1}=0$ to provide a reference point (i.e., a carrier phase attached to the envelope $A(t)$); the {\it relative} spectral phase controls $\{\phi_m\}$ are with respect to this reference. With consideration of the transition frequencies in Eq. (\ref{ho}), the $M+1$ field frequencies are $\omega_m=m$, $m=1,2,3,\ldots M+1$ and $T=30$. A fixed Gaussian envelope $A(t)=\sqrt{F}\thinspace\text{exp}[\frac{8\pi}{T^2}(t-\frac{T}{2})^2]$ is employed to ensure smooth switching on and off of the control field. The integrated field fluence is $F$, where the simulations employ values of $F$ ranging from 0.069 to 10. By construction, the fluence remains fixed throughout an optimization search over the phases $\{\phi_m\}$ because $A(t)$ is not allowed to vary. In order to ensure a sufficiently fine time-resolution such that no false traps arise on the landscape due to the additional constraint of time-discretization \cite{mepif}, we utilized 1024 time-points for $M<6$, 2048 time-points for $6\leq M\leq 12$, and 4096 time-points for $M\geq13$.

\subsection{Optimization procedure for climbing the landscape}\label{alg}
This work employs a gradient search procedure to determine the change in the controls $\{\phi_m\}$ at each algorithmic step because it is sensitive to the landscape topology, meaning that the algorithm will halt upon reaching a suboptimal trap on the control landscape. For implementation of the gradient search, the controls $\{\phi_m\}$ may be written as a vector $\Phi=[\phi_1,\thinspace\phi_2,\ldots\phi_M]$. We define a variable $s$ specifying the progress of the optimal search from an initial random vector $\Phi(s=0)$ to a vector specifying a critical point $\Phi^{\star}=\Phi(s=S)$. $S$ is the value of $s$ corresponding to a landscape point that satisfies the critical point condition $\frac{\partial \mpif}{\partial\Phi^{\star}}\simeq\mathbf{0}$, where $\mathbf{0}$ denotes the zero vector. The landscape value $\mpif(s)\equiv \mpif[\Phi(s)]$ depends upon $s$ through the dependence of $\Phi(s)$ on $s$. Thus, a differential change in the landscape value $d\mpif$ associated with a differential change $ds$ is given by $d\mpif\equiv \left(\frac{\partial \mpif}{\partial s}\right)ds$ and the chain rule,

\begin{equation}
\frac{d\mpif}{ds}\equiv \frac{\partial\mpif}{\partial\Phi(s)}\frac{\partial \Phi(s)}{\partial s}.\label{dJds}
\end{equation} 
As the objective is to maximize $\mpif$, we have the demand that $\frac{d\mpif}{ds}\geq0$, which specifies that $\Phi(s)$ must satisfy the differential equation 
\begin{equation}
\frac{\partial\Phi(s)}{\partial s}=\gamma\frac{\partial\mpif}{\partial\Phi(s)},\quad \gamma>0.\label{ode}
\end{equation}
The analytical expression for the gradient on the right-hand side of Eq. (\ref{ode}) will be derived below in Section \ref{gradhess}. The present search algorithm, incorporated into MATLAB (routine \texttt{ode45}) \cite{matlab}, solves Eq. (\ref{ode}) using a fourth order Runge-Kutta integrator with a variable step size $\gamma$ to determine $\Phi$ at the next iteration. The search process is terminated when either (a) the \pif\ms value reaches the desired convergence criterion (\pif$>0.999$ in Sections \ref{num} and \ref{exp}, and \pif$>0.999999$ in Section \ref{optlset}) or (b) the \pif\ms value between consecutive iterations increases by less than $10^{-8}$. The latter situation (b) indicates that a suboptimal critical point has been reached.

\subsection{Procedure for exploring optimal level sets}\label{lset}
Analytical and numerical evidence \cite{demiralp,mike2,vinny2} shows that the critical value $\mpif=1$ lies on a submanifold locally corresponding to the eigenvectors of the Hessian matrix with associated null eigenvalues. If the number of controls $M$ satisfies $M>2N-2$ and they are sufficient to achieve $\mpif=1$, then it is expected that the Hessian will contain at least $M-2N+2$ null eigenvalues and corresponding eigenvectors. The resulting optimal level set of connected controls may be explored by continuously varying the controls such that the optimal condition $\mpif=1$ is maintained \cite{vinny2}. This movement on the level set requires satisfaction of the second order optimality condition
\begin{align}
\frac{d^2\mpif}{ds^2}&=\sum_{m}\sum_{\mprime}\frac{\partial \phi_m(s)}{\partial s}\mathcal{H}(\phi_m(s),\phi_{\mprime}(s))\frac{\partial \phi_{\mprime}(s)}{\partial s}=0.\label{2o}
\end{align}
The Hessian matrix $\mathcal{H}(\phi_m,\phi_{\mprime})=\frac{\partial^2\mpif}{\partial\phi_m\partial\phi_{\mprime}}$ corresponds to the second-order variation of \pif\ms with respect to the control phases $\phi_m$ and $\phi_\mprime$. An analytical expression for the Hessian will be derived in Section \ref{gradhess}. To ensure that Eq. (\ref{2o}) holds, one may only move the controls $\{\phi_m(s)\}$ in the direction of the eigenvector(s) of the Hessian with associated null eigenvalues. 

This work will consider the illustrative case at $\mpif=1$ with $M=2N-1$ in Section \ref{lsetexp}. The choice of $M=2N-1$ places the number of controls just above the threshold of $2N-2$, considered as the generally minimum number needed to reach $\mpif=1$ \cite{mike2,me}. Although more than one null Hessian eigenvalue can exist when $M=2N-1$ \cite{vinny2}, all of the optimal level sets found in this work were one-dimensional when $M=2N-1$. The impact of operating with other values of $M$ on the landscape topology will be assessed in Section \ref{num}. In the case of $M=2N-1$, the appropriate direction of movement on the level set at $\mpif=1$ to satisfy Eq. (\ref{2o}) is
\begin{equation}
\frac{\partial{\Phi}}{\partial s}=\pm Q_0(s)\label{lo}
\end{equation}
where $Q_0$ is the Hessian eigenvector corresponding to the zero eigenvalue. The level set may be mapped out by taking small steps in $s$ along the direction of $\pm Q_0(s)$, followed by a recalculation of the Hessian and an updated null eigenvector as $s$ evolves. Since numerical inaccuracies inevitably result in some variation of \pif\ms with increasing $s$, the optimal value $\mpif\simeq1$ is preserved by alternating level set exploration steps (i.e., Eq. (\ref{lo})) with gradient climbs via Eq. (\ref{ode}) when the \pif\ms value falls outside a specified range. In the simulations, when an optimal level set at $\mpif=0.999999$ is being explored and the \pif\ms value drops below $\mpif=0.99999$, a gradient climb is employed to regain the value $\mpif=0.999999$, and the level set exploration is continued from this new point. This automatic correction procedure was found to be adequate to ensure faithful exploration of optimal level sets.

As explained above, controlling the quantum dynamics to climb the landscape and traverse an optimal level set calls for solving Eqs. (\ref{ode}) and (\ref{lo}), respectively. The right-hand side of these differential equations are highly non-linear in the controls $\{\phi_m\}$ through the time evolution operator $U(t,0)$ (c.f., Section \ref{gradhess}). The treatment of the landscape exploration dynamical equations (\ref{ode}) and (\ref{lo}) can be considered in analogy to the behavior of non-linear dynamical systems \cite{guckenheimer,strogatz}. In this regard, the optimal level sets are functions of periodic controls, i.e., the control values $\phi_m$ and $\phi_m\pm 2\pi n_m$ for any integer $n_m$ produce identical control fields $\e(t)$ through Eq. (\ref{et}). Thus, it is possible that the level sets may form periodic trajectories over $s$ in the control space $\{\phi_m\}$. Such periodic level sets are analogous to {\it periodic orbits} in nonlinear dynamical systems \cite{guckenheimer,strogatz}. However, the optimal level sets do not necessarily have to be periodic. In this work, we will show that both periodic level sets and aperiodic `wandering' level sets are present on the constrained top of the landscape. The distinct nature of periodic orbits and `wandering sets' with no periodic structure is well-documented in the non-linear dynamics literature \cite{guckenheimer,strogatz}. 

The `size' of a periodic level set can be measured by the total path length $\cal L$ traversed by $\{\phi_m\}$ in $s$ over one period,  which may be computed by the integral
\begin{equation}
{\cal L}=\int_0^{s^*} ds \left[\sum_m\left(\frac{d\phi_m(s)}{ds}\right)^2 \right]^{1/2}\label{pathl}
\end{equation}
on the closed trajectory defined from $s=0$ to $s=s^*$. At the point $s^*$, {\it each} of the $\phi_m$ is shifted by an integer $n_m$ multiple of 2$\pi$ from its value at $s=0$. This shift may be zero, with the phases $\{\phi_m\}$ at $s^*$ returning to their initial values. Alternatively, one or more of the $\phi_m$ may grow or decrease in magnitude by $2\pi n_m$ over a period and still return to a physically equivalent point in control space, i.e., producing an identical control field $\e(t)$ through Eq. (\ref{et}) while maintaining $\mpif\simeq1.0$. In Eq. (\ref{pathl}), the derivative $d\phi_m(s)/ds$ between the steps $s_i$ and $s_{i+1}$ is approximated as $d\phi_m(s)/ds=(\phi_m(s_{i+1})-\phi_m(s_i))/(s_{i+1}-s_i)$. The path length $\cal L$ provides a quantitative metric of the extent of periodic level sets in the control space and will be calculated for the optimal level sets in Section \ref{lsetexp}. For `wandering' level sets that do not have a periodic structure, the path length would continue to grow as $s$ increases. This circumstance can include cases where individual phases $\phi_m$ pass through $\phi_m\pm2\pi n_m$, but at distinct values of $s$ for each phase $m$. The numerical evidence for the existence of both types of optimal level sets will be assessed in Section \ref{lsetexp}.

\subsection{Derivation of the Gradient and Hessian of \pif\ms with respect to the controls}\label{gradhess}
A practical expression for the first-order variation of \pif\ms with respect to the controls on the right-hand side of Eq. (\ref{ode}) is needed to implement the gradient search procedure in Section \ref{alg}. The procedure for exploring optimal level sets in Section \ref{lset} requires an expression for the Hessian matrix, or the second-order variation of \pif\ms with respect to the controls, $\frac{\partial^2\mpif}{\partial\phi_m\partial\phi_{\mprime}}$. The analysis below derives these expressions. 

The variation of $P_{i\to f}$ due to functional changes in the Hamiltonian $\delta H(t)$ is obtained by considering the resultant responses $\delta U(t,0)$:
\begin{align}
&i \frac{\partial}{\partial t}\delta U(t,0)=H(t)\delta U(t,0)+\delta H(t)U(t,0),\qquad \delta U(0,0)=0\label{du}\\
&\delta P_{i\to f}=\langle i|\delta U^{\dag}(T,0)|f\rangle\langle f|U(T,0)|i\rangle+\langle i|U^{\dag}(T,0)|f\rangle\langle f|\delta U(T,0)|i\rangle\label{dp}.
\end{align}
Eq. (\ref{du}) can be integrated \cite{taksan} to give
\begin{equation}
\delta U(t,0)=-{i}\int_0^t U(t,t')\delta H(t')U(t',0)dt',\label{intdu}
\end{equation}
and substitution of Eq. (\ref{intdu}) into Eq. (\ref{dp}) gives
\begin{equation}
\delta P_{i\to f}=-2{\rm Im} \int_0^T\langle i|U^{\dag}(T,0)|f\rangle\langle f|U(T,0)U^{\dag}(t,0)\delta H(t) U(t,0)|i\rangle dt\label{intdp}.
\end{equation}
Within the dipole formulation, $\delta H(t)=-\mu\delta\e(t)$, the functional derivative $\delta P_{i\to f}/\delta \e(t)$ becomes
\begin{align}
\frac{\delta P_{i\to f}}{\delta\e(t)}=-2\textup{Im}\left[\langle i|U(T,0)\dg|f\rangle\langle f|U(T,0)U\dg(t,0)\mu U(t,0)|i\rangle\right].\label{grad}
\end{align}
From Eq. (\ref{grad}), the first derivative of \pif\ms with respect to the controls $\{\phi_m\}$ may be obtained by the chain rule using the expression for the control field in Eq. (\ref{et}),
\begin{align}
\frac{\partial\mpif}{\partial\phi_m}&=\int_0^T\frac{\delta\mpif}{\delta\e(t)}\frac{\partial\e(t)}{\partial\phi_m}dt=2\ms\text{Im}\int_0^T\langle q|k\rangle A(t)\sin(\omega_mt+\phi_m)dt,\label{dpdphi}
\end{align}
where $|q\rangle=U^{\dag}(T,0)|f\rangle\langle f|U(T,0)|i\rangle$ and $|k\rangle=U\dg(t,0)\mu U(t,0)|i\rangle$. 

The Hessian matrix $\mathcal{H}(\phi_m,\phi_{\mprime})=\frac{\partial^2\mpif}{\partial\phi_m\partial\phi_{\mprime}}$ is obtained from differentiation of Eq. (\ref{dpdphi}) with respect to the control $\phi_{\mprime}$:
\begin{align}
\frac{\partial^2\mpif}{\partial\phi_m\partial\phi_{\mprime}}=&2\thinspace\text{Im}\int_0^T\bigg(\bigg[\langle i|\frac{\partial U^{\dag}(T,0)}{\partial\phi_{\mprime}}|f\rangle\langle f|U(T,0)|k\rangle+\langle i|U\dg(T,0)|f\rangle\langle f|\frac{\partial U(T,0)}{\partial\phi_{\mprime}}|k\rangle\nonumber\\
&+\langle q|\frac{\partial U^{\dag}(t,0)}{\partial\phi_{\mprime}}\mu U(t,0)|i\rangle
+\langle q|U\dg(t,0)\mu\frac{\partial U^{\dag}(t,0)}{\partial\phi_{\mprime}}|i\rangle\bigg]A(t)\sin(\omega_mt+\phi_m)\nonumber\\
&+\delta(m,\mprime)\langle q|k\rangle A(t)\cos(\omega_mt+\phi_m)\bigg) dt,\label{delder2}
\end{align} 
where $\delta(m,\mprime)$ denotes the Kronecker delta function. The derivatives $\partial U(t,0)/\partial\phi_{\mprime}$ and $\partial U(T,0)/\partial\phi_\mprime$ are computed with Eq. (\ref{intdu}) for $\delta H=-\mu\delta\e(t)$ along with Eq. (\ref{grad}) in a fashion analogous to the procedure in Eq. (\ref{dpdphi}). Using the shorthand notation $\mu(t)=U\dg(t,0)\mu U(t,0)$, Eq. (\ref{delder2}) simplifies to
\begin{align}
\mathcal{H}(\phi_m,\phi_{\mprime})&=2\thinspace\text{Re}\int_0^Tdt\bigg[-\langle i|\int_0^Td\tprime\mu(\tprime)A(\tprime)\sin(\omega_{\mprime}\tprime+\phi_{\mprime}) U^{\dag}(T,0)|f\rangle\langle f|U(T,0)|k\rangle\nonumber\\
&+\langle q|\int_t^Td\tprime\mu(\tprime)A(\tprime)\sin(\omega_{\mprime}\tprime+\phi_{\mprime})|k\rangle\nonumber\\
&+\langle q|\mu(t)\int_0^td\tprime\mu(\tprime)A(\tprime)\sin(\omega_{\mprime}\tprime+\phi_{\mprime})|i\rangle\bigg] A(t)\sin(\omega_mt+\phi_m)\nonumber\\
&+2\ms\text{Im}\int_0^Tdt\ms\delta(m,\mprime)\langle q|k\rangle A(t)\cos(\omega_mt+\phi_m).\label{hess}
\end{align}
The Hessian is the $M\times M$ dimensional matrix whose elements are given by Eq. (\ref{hess}), which is valid anywhere on the landscape. Of particular importance are the Hessian eigenvalues and eigenvectors at the critical points on the landscape, as these are needed for determining the local topology and for exploration of optimal level sets of controls producing $\mpif\simeq1.0$, as described in Section \ref{lset}. The Hessian eigenvalue spectrum at suboptimal critical points on the landscape is also important because it can be used to classify these landscape points as traps or saddles. 

\section{Assessing the prevalence and location of traps on constrained control landscapes}\label{num}
Any evaluation of the effects of the number of independent controls on the landscape topology must address the nature of the controls being employed. The spectral phase controls $\{\phi_m\}$ utilized in this work may produce distinct landscapes with different choices of spectral frequencies; for example, a poor choice would have all of the frequencies $\{\omega_m\}$ far away from the $H_0$ transition frequencies. Here, we reasonably choose the frequencies to be within the bandwidth of the $H_0$ transitions. The equal spacing of the frequencies $\omega_m=m,\ms m=1,2,\ldots M+1$ results in one or more of the frequencies being resonant with the transitions in $H_0$ when $m$ is even, (depending on $N$, c.f., Eq. (\ref{ho})), and the remaining frequencies (e.g., for odd $m$) are included in order to take advantage of potential Stark shifting of the energy levels upon interaction with the control field. All of the frequencies $\omega_m$ are shifted to be off-resonant with the $H_0$ transitions in one illustrative case as well. Although the numerical simulations consider only the constrained field in Eq. (\ref{et}), the landscape exploration concepts and tools presented here can readily be applied to any other choice of constrained field. It is anticipated that many such studies will be needed to fully assess the impacts of constrained fields.

In order to determine the prevalence of traps on the landscape, we employ the local gradient algorithm discussed in Section \ref{alg} because it halts upon reaching a suboptimal critical point. The simulations consider the number $M$ of controls that are required to (a) enable at least one search out of 100 to achieve $\mpif=1$, or (b) ensure that $each$ search achieves the optimal value of \pif. Satisfaction of (a) indicates that few enough constraints are present to ensure the existence of at least one optimal point on the constrained landscape, and satisfaction of (b) suggests that $M$ is sufficiently large to eliminate false traps on the landscape. In between these extremes, the goal is to assess both the probability ${\cal P}_c$ of reaching $\mpif=1$ as a function of $M$, as well as the location of any observed traps on the landscape (i.e., in terms of their \pif\ms values). Even when traps are present, their effect is less detrimental to optimization when they occur at high \pif\ms values near the optimum, as closely approaching $\mpif=1.0$ is not often necessary in practical applications. As explained in Section \ref{meth}, an analysis of the gradient and Hessian of \pif\ms with respect to the controls predicts that at least $2N-2$ well-chosen controls may be necessary to expect a high probability of convergence \cite{demiralp,mike2}. This conjecture was found to be valid when using kinematic controls (e.g., the matrix elements of the propagator $U$), which are independent of the Hamiltonian \cite{me}. Furthermore, since only $N^2$ well-chosen independent controls are required to generate $U$, it is expected that choosing $M=N^2$ will be sufficient to ensure convergence to an optimal value of \pif\ms without encountering traps on the control landscape, provided that no other control constraints (e.g., limited field fluence) are present. 

Systems of $N$ states ranging from 3 through 8 using Eqs. (\ref{ho}) and (\ref{mu}) were employed with control fields of the form in Eq. (\ref{et}). Each initial phase $\phi_m(s=0)$ was selected from a uniform distribution on $[0,2\pi]$ and allowed to vary freely as a function of $s$ during optimization. Field fluence values of $F=10$, $F=1$, and $F=0.1$ were employed, where $F$ remains fixed throughout each optimization via Eq. (\ref{et}). The $|1\rangle\to|N\rangle$ target transition was chosen for optimization because it was found to be the most difficult case to optimize, especially as $N$ rises, and thus presents the most stringent test of the control landscape topology \cite{mepif}. The convergence criterion was $P_{1\to N}>0.999$ and the number of controls $M$ in Eq. (\ref{et}) ranged from $2N-5$ through $N^2$. A total of 100 searches starting at random values of the phases were run for each choice of $N$, $M$, and $F$ to provide convergence statistics. 

\subsection{Convergence probability}\label{n1}
The prevalence of false traps on the control landscape is assessed by the probability ${\cal P}_c$ of converging to the optimal value $\mpin\geq0.999$, where a unit convergence probability ${\cal P}_c=1$ suggests a lack of traps on the landscape. The convergence probability ${\cal P}_c$ of reaching $\mpin\geq0.999$ at selected values of $N$, $M$ and $F$ is plotted in Figure \ref{convstat}, where the abscissa values $M$ are given in terms of $N$ (e.g., $M=2N-2$) in order to show the dependence of the convergence probability ${\cal P}_c$ on the number of controls $M$ relative to $N$.

For $F=10$ (the circles in Figure \ref{convstat}), the convergence probability ${\cal P}_c$ at each value of $M$ was found to vary little when $N$ was changed, so the recorded value of ${\cal P}_c$ is averaged over all $N$ from 3 through 8. At least one search was found to reach $\mpin\geq0.999$ when $M\geq2N-4$, and ${\cal P}_c=1$ was achieved when $M\geq2N+1$. In order to confirm that $M=2N+1$ is sufficient to ensure convergence without encountering traps, an additional 1000 optimal searches were performed for $N=3$ with $M=7$ and $N=8$ with $M=17$. One search became trapped at $P_{1\to3}\sim0.997$ for $N=3$ and two searches were trapped at $P_{1\to8}\sim0.998$ for $N=8$. Thus, traps exist on the landscape when $M=2N+1$ with $F=10$, but they are extremely rare and occur very close to the landscape top. No traps were found using $M=2N+2$ in an additional 1000 searches for $N=3$, indicating a high likelyhood of a trap-free landscape topology within this constrained form of the control field for sufficiently large $M$. 

Unlike the case of $F=10$, the convergence probability ${\cal P}_c$ at a fixed $M$ does depend on $N$ for $F=1$, as shown by the distinct patterns of convergence exhibited by $N=3$, 5, and 8 in Figure \ref{convstat} (the squares, triangles, and diamonds, respectively). For $N\leq 5$, a total of $M\geq2N-4$ controls are necessary to attain $\mpin\geq0.999$ (i.e., at least one search out of 100 reaching it), and $M\geq2N-3$ is needed for larger $N$. The value of $M$ required to achieve ${\cal P}_c=1$ grows with $N$, from $M\geq2N$ when $N=3$ and 4 to $M>2N+2$ for $N=8$ (i.e., at $N=8$, ${\cal P}_c=0.97$ with $M=2N+2$). The observed decrease in ${\cal P}_c$ as $N$ rises is due to the difficulty of optimizing the $|1\rangle\to|N\rangle$ transition as $N$ increases \cite{mepif}, and demonstrates that beyond limiting the number of controls, the fluence imposes an additional significant constraint on optimization. For $F=0.1$ and $N=3$, a few searches did not converge even upon choosing $M=N^2$, indicating that the fluence is imposing a significant additional constraint since $N^2$ well-chosen controls should be sufficient to generate any propagator $U(T,0)$. Because of this strong additional constraint, searches with $N>3$ were not performed with $F=0.1$.

For all values of $F$, $M=2N-2$ controls are needed to achieve ${\cal P}_c\simeq0.5$, while using fewer controls drastically reduces the convergence probability to ${\cal P}_c<0.2$. It is also apparent from Figure \ref{convstat} that $M=2N-2$ corresponds to the greatest rate of change in the value of ${\cal P}_c$, in agreement with observations using kinematic controls \cite{me} and corresponding with analysis of the gradient and Hessian with respect to \pif\ms \cite{demiralp,mike2}. However, the `rule' that $2N-2$ controls are sufficient to produce ${\cal P}_c\simeq0.5$ still requires a good physically relevant choice of controls. In many practical situations, the best choice of controls may not be known {\it a priori}, and the value of $N$ may be unknown as well for experimental objectives such as molecular fragmentation \cite{Levis2001, Gerber02,Gerber2002, Weinacht2005b,melab2}. In these situations, the minimal necessary $M$ and the best choice of controls must be found by performing experiments \cite{Weinacht2005b,melab2}.

As an illustration of the importance of choosing a good set of $M=2N-2$ controls, we assess the effects of shifting the $M+1$ frequencies of $\e(t)$ in Eq. (\ref{et}) progressively further away from the transitions in $H_0$ for the illustrative case of $N=3$, $M=4$, and $F=10$. Optimization searches were performed for the sets of $M+1=5$ frequencies $\omega_m=m+\Omega$ in Eq. (\ref{et}) for frequency shift values from $\Omega=3$ through $\Omega=19$. For comparison, the maximal transition frequency in $H_0$ is $H_0(3,3)-H_0(1,1)=6$, so all of the field frequencies shift to higher values than the $H_0$ transitions when $\Omega$ is sufficiently large. 100 searches seeking optimal values of the controls $\{\phi_m\}$ were performed for each set of frequencies defined by $\Omega$. The convergence probability ${\cal P}_c$ (blue squares), maximal $P_{1\to 3}$ value (red circles), and mean value $\langle P_{1\to 3}\rangle$ (green triangles) as a function of $\Omega$ are plotted in Figure \ref{omega}, where the value $\Omega=5$ is shown as the dashed vertical line, corresponding to the lowest spectral frequency being equal to $H_0(3,3)-H_0(1,1)=6$. When $\Omega+1>6$, the value of ${\cal P}_c$ significantly decreases. Further increasing $\Omega$ results in a decrease in the mean yield $\langle P_{1\to 3}\rangle$, and when $\Omega\geq14$, no searches reached the optimum $P_{1\to 3}>0.999$. These results show that the `$2N-2$ rule' can only be expected to hold for a well-chosen form of the control field (here, where the frequencies of the control field overlap the transitions in $H_0$). Thus, the specific nature of limited controls can have an impact on the {\it apparent} topology of the control landscape. 

\subsection{Location of constraint-induced traps on the landscape}\label{n2}
Further information about the location of false traps (i.e., expressed in terms of \pif\ms value) on the landscape can be obtained from examining the statistical mean $\langle\mpif\rangle$ over the sample of 100 searches as a function of $N$ and $M$. Because a high-quality yield (e.g., $\mpif>0.999$) is not necessary for many practical applications, assessing the location of traps on the landscape in terms of their \pif\ms values is important. If traps exist predominantly at high \pif\ms values, e.g. $\mpif>0.95$, these may not preclude convergence to an acceptable yield in practice. In contrast, the existence of constraint-induced traps at significantly lower yields could pose a greater impediment to identifying control fields that produce a suitably high yield. In such cases, a stochastic search algorithm may overcome such traps to some degree if a high yield is still accessible with the field constraints present.

The mean value $\langle\mpin\rangle$ over 100 searches is plotted as a surface versus $N$ and $M$ for $F=10$ and $F=1$ in Figures \ref{d3}(a) and \ref{d3}(b), respectively. For both values of $F$, employing $M>2N-2$ typically results in $\langle\mpin\rangle\geq 0.99$, further indicating that choosing more than $2N-2$ judicious controls may be sufficient for practical optimization applications. The distinct shapes of the resulting surfaces show that choosing $F=10$ imposes no further constraint on the optimization beyond the number of controls, but $F=1$ is small enough to induce a further constraint. For $F=10$, $\langle\mpin\rangle$ rises with $N$ for the same value of $M$. This behavior is intuitive if no other constraints are present, as the relative difference between adjacent values of $M$ decreases with rising $N$ (e.g., consider $M=2N-2$ and $M=2N-3$, which is $M=14$ versus $M=13$ for $N=8$, but the corresponding values of $M=4$ versus $M=3$ for $N=3$), so the detrimental effect of reducing $M$ from $2N-2$ to $2N-3$ is expected to be smaller for $N=8$ than for $N=3$. In contrast, at $F=1$, increasing $N$ results in a decrease of $\langle\mpin\rangle$ for fixed $M$ in some cases. This distinct behavior arises because the control energy required to reach the $|1\rangle\to|N\rangle$ transition increases with $N$ \cite{mepif}, and indicates that the small fluence $F=1$ imposes an additional constraint.

\section{Exploring suboptimal critical regions on the landscape}\label{exp}
The previous section examined the statistical prevalence and location of traps on constrained control landscapes. This section explores the topology and features of suboptimal critical regions for selected illustrative cases. This information is important for identifying optimization procedures, including advanced stochastic algorithms, that may avoid or mitigate their effects on the attainable optimization yield and/or the convergence rate. 

\subsection{Constraint-induced saddles}
The results of Section \ref{num} show that traps arise with inadequately flexible controls, reflected in the gradient climbing algorithm getting `stuck' at a suboptimal value of \pif. The appearance of traps also raises the question of whether saddle regions may arise on constrained control landscapes, even though no saddles are present on the unconstrained \pif\ms landscape \cite{mike1,demiralp}. Saddle points on the landscape satisfy $\partial \mpif/\partial\Phi=\mathbf{0}$, but have an indefinite Hessian spectrum that contains a combination of positive, negative, and null eigenvalues. While a trap has a negative semi-definite Hessian, meaning that the \pif\ms value cannot increase upon making $any$ small perturbation in the controls, there is at least one positive Hessian eigenvalue leading off of a saddle point. Thus, a small perturbation in the controls in the direction of the corresponding Hessian eigenvector could improve the \pif\ms value and escape from a saddle point. However, gradient searches may be attracted by saddles and significantly slow down in the vicinity of a saddle point. This effect has been shown to increase the search effort in the cases that encounter saddle points for unconstrained optimization of objectives whose landscapes inherently contain these features \cite{greg, mew}. 

Some searches reaching $\mpin\geq0.999$ in Section \ref{num} display evidence of encountering one or more saddles, as measured by the flattening of the trajectory $\mpin(s)$ versus $s$ at a sub-optimal value of $\mpin<0.999$ before ultimately reaching the optimum. One example of a search with $N=3$, $M=4$, and $F=1$ that encountered a saddle at $P_{1\to 3}=0.9$ is shown in Figure \ref{saddlef}. The top panel (a) shows the $P_{1\to 3}$ value versus $s$ (blue trajectory), which flattens at $P_{1\to 3}=0.9$ from $s\simeq0.1$ to $s\simeq2$. For comparison, a trajectory of $P_{1\to 3}$ versus $s$ for a search that did not encounter a saddle is shown in green. Examination of the Hessian eigenvalues in Figure \ref{saddlef}(b) confirms the presence of a saddle point along the blue trajectory of Figure \ref{saddlef}(a); in this region three eigenvalues are distinctly negative, while the fourth has a very small positive value of 0.003 at $s=0.91$. For reference, the negative eigenvalues typically fall between the values -0.05 and -10 at a local or global maximum when $N=3$, $M=4$, and $F=1$. Starting at $s\simeq2$, the search clearly begins to escape the saddle, as shown by the increase in the rate of $P_{1\to 3}$ climbing in Figure \ref{saddlef}(a), until the search reaches the optimum. A total of 14 searches of the 68 that converged to $P_{1\to 3}\geq0.999$ for $N=3$, $M=4$, $F=1$ were found to encounter saddles through analysis of the $P_{1\to 3}$ trajectories, the gradient norm $||\partial P_{1\to 3}/\partial \Phi||$, and the Hessian eigenvalues. At each of these saddles, the gradient norm decreases to less than 0.003 while in the vicinity of the saddle; for reference, the gradient norm at $P_{1\to3}\approx0.999$ ranges from 0.002 to 0.005 for these searches.  At each saddle, only one of the Hessian eigenvalues is positive, with values ranging from 0.002 to 0.1. In contrast, the largest magnitude negative eigenvalue in these regions is less than -10, indicating that the search direction leading off of these saddles is very flat. This feature results in the search trajectories significantly slowing down, spending between 20$\%$ and 60$\%$ of the entire search trajectory (as measured by $s$) in the vicinity of the saddle point. This behavior is evident in Figure \ref{saddlef}(a), where $\sim40\%$ of the search trajectory is spent around the saddle value. Overall, approximately $10\%-30\%$ of searches at a given $N$ and $F$ for $M$ between $2N-2$ and 2$N$ that converge to $\mpin\geq0.999$ appear to encounter saddles based on examination of their $\mpin$ trajectories as a function of $s$, typically at $\mpin\gtrsim0.9$. The presence of saddles in cases where few or no traps exist (i.e., for $M\geq2N-2$) suggests that imposing mild constraints may introduce saddles on the landscape, while more severe constraints turn these saddles into trapping extrema. 

\subsection{Constraint-induced traps}\label{sub}
Although encountering a saddle typically increases the effort required to find an optimal solution, their presence on the landscape does not prevent successful optimization. In contrast, encountering a sub-optimal trap will halt a gradient and possibly other local search algorithms. The prevalence of traps on the control landscape was assessed in Section \ref{num}; here we examine their topology through analysis of the Hessian eigenvalues. The vast majority of traps observed have negative definite Hessians with the smallest negative Hessian eigenvalue typically $\lesssim-0.05$ and the largest negative Hessian eigenvalue typically $\sim-10$, and are thus {\it isolated} points. All traps observed when employing $M\leq2N-2$ were found to be isolated points. However, in rare cases when $M\geq2N-1$, one  `zero' Hessian eigenvalue between -0.002 and 0.005 was observed, which is approximately $0.05\%$ of the magnitude of the largest negative Hessian eigenvalue and an order of magnitude smaller than the typical smallest negative Hessian eigenvalue. For reference, eight of the 14 saddles observed for $N=3$, $M=4$, and $F=1$ had one positive eigenvalue of less than 0.01, with the remaining eigenvalues being less than -0.05. Thus, the `zero' Hessian eigenvalues examined here are of similar magnitude to the lone small positive Hessian eigenvalue encountered at some saddle points.

Representative examples of putative traps with one `zero' Hessian eigenvalue were further examined in order to ascertain the true topology of the landscape in these regions. The second-order search method described Section \ref{lset} was employed to examine these landscape points, with the search trajectory in Eq. (\ref{lo}) directed along the Hessian eigenvector with the `zero' eigenvalue. In order to ensure that the search trajectory does not `fall off' the original value  $\mpin^\star$, where the star $\star$ denotes the trapped $\mpin$ value, the direction of the `zero' eigenvalue was followed as long as the $\mpin$ value remained above $\mpin^\star-0.0001$; when the $\mpin$ value fell below this threshold, the gradient climbing method of Section \ref{alg} was employed until the $\mpin$ value improved by less than $10^{-8}$ at consecutive $s$-steps (i.e., the trapping criterion in Section \ref{alg}). In some cases, this procedure resulted in a $\mpin$ value greater than the initial value $\mpin^\star$.

Overall, 17 alleged traps containing one `zero' eigenvalue were investigated for $N$ ranging from 3 to 6 and $M$ from $2N-1$ to $2N+2$. Eight of these attained an optimal value $\mpin\geq0.999$ upon moving in the direction of the eigenvector associated with the `zero' eigenvalue. Some cases also required a gradient climb, as discussed above. This behavior indicates that the putative trap is, in fact, a saddle feature, because it can be escaped upon moving in the direction corresponding to the smallest magnitude Hessian eigenvalue. The saddle behavior is presented for five search trajectories of $\mpin$ versus $s$ beginning with apparent traps at $\mpin^\star$ values ranging from 0.986 to 0.998 in Figure \ref{trapsaddle}. The cases for $N=3$ and $N=4$ shown in the figure attained $\mpin\geq0.999$ solely by moving in the direction of the Hessian eigenvector corresponding to the `zero' eigenvalue, while the two cases for $N=6$ attained optimal $\mpin\geq0.999$ after employing the gradient climb when movement in the direction of the `zero' eigenvalue no longer increased the $\mpin$ value. The remaining nine cases investigated were found to be isolated trapping points, with $\mpin$ values decreasing from the initial value $\mpin^\star$ after few steps along the `zero' eigenvalue direction, and subsequent gradient climbs not reaching the original value $\mpin^\star$. The saddle topology of some apparent traps with `zero' eigenvalues is consistent with the observed saddle topology for some of the $N=3$, $M=4$, $F=1$ searches, with the Hessian direction leading off of the saddle corresponding to a very small magnitude eigenvalue. This prevalence of saddles on the landscape, some of which were found to halt the gradient search algorithm employed here, indicates that a second-order procedure (e.g., conjugate gradient search) may in some cases be superior for searching over constrained landscapes, as searches encountering saddles would likely escape from them more quickly.

\section{Exploring optimal regions at the top of the landscape}\label{optlset}
In order to adequately examine the topology at optimal controls on the top of the field-constrained landscape, searches from Section \ref{num} that converged to $\mpif\geq0.999$ were further optimized to $\mpif\geq0.999999$. This more stringent convergence criterion ensures operation close enough to the top of the landscape such that the predicted topology of optimal regions can be examined numerically \cite{demiralp,vinny2}. In particular, the predicted number of $N^2-2N+2$ null Hessian eigenvalues was confirmed, with each converging to a very small value above $-0.002$, which is $\sim0.02\%$ of the magnitude of the largest negative Hessian eigenvalue.

\subsection{Hessian eigenvalue spectra for constrained optimal fields}\label{heig}
Optimization to the yield $\mpif\geq0.999999$ was performed for $N=3$ with $M=4$ through 7, $N=4$ with $M=6$ through 8, and $N=5$ with $M=8$ and 9. For a given choice of $N$, $M$, and $F$, approximately $10\%-40\%$ of the searches from Section \ref{num} that converged to $\mpif=0.999$ became trapped below $\mpif=0.999999$ (likely due to numerical discretization upon solving the Schr\"odinger equation acting as a further constraint at high yields \cite{mepif}) and were discarded for the following analysis. The mean of each Hessian eigenvalue over the subset of the 100 searches that converged to $\mpif\geq0.999999$ (at least 30 searches in each case) is plotted versus its index in Figure \ref{eigs} for $F=10$; the results when $F=1$ are similar (not shown). The left and right standard deviations from the mean are shown by the error bars for the representative case of $N=4$, $M=8$; statistical distributions for the remaining $N$ and $M$ values were similar. For all $N=3$ searches, there is a clear break between eigenvalues 4 and 5, corresponding to eigenvalues labeled by $2N-2$ and $2N-1$, respectively. A similar trend is observed for $N=4$ between eigenvalues 6 and 7, while for $N=5$, the corresponding jump occurs between eigenvalues 8 and 9. This behavior confirms the predicted features of the Hessian spectrum at the top of the landscape \cite{demiralp}. In all of the cases, no `zero' eigenvalues were found when $M=2N-2$, indicating that these optimal solutions are isolated points. This circumstance can be interpreted in terms of the underlying optimal submanifold `shrinking' to an isolated point as the number of controls is decreased to the critical value of $2N-2$. In these cases, increasing $M$ to $2N-1$ produces a single `zero' Hessian eigenvalue corresponding to the existence of a one-dimensional level set at the top of the landscape.

\subsection{Optimal level sets}\label{lsetexp}
Picking up on the last comment above, we explore one-dimensional optimal level sets for illustrative cases with $N=3$ and $M=2N-1=5$ using the second-order search method described in Section \ref{lset}. Search trajectories are initiated from an optimal point on the top of the landscape identified from the studies in Section \ref{heig}. A total of five level sets for $F=1$, six level sets for $F=0.1$, and multiple level sets for $F<0.1$ were explored. Additional level sets likely exist on each landscape beyond those examined here, and similar features are expected to arise for those as well. These observed level set features for the simple case of phase controls with $N=3$, $M=5$ provide an illustration of behavior that may be encountered upon operating with any type of constrained control field beyond the formulation in this work. As discussed in Section \ref{lset}, optimal level sets may be aperiodic `wandering' sets or have a periodic structure in control space. 

Three level sets that appeared to be wandering were discovered on the landscape with $N=3$, $M=5$, and $F=1$, listed in Table \ref{lst}. The trajectories of these three level sets never returned to the same point in control space, even allowing for $2\pi$ shifts in the controls $\{\phi_m\}$, which strongly suggests that they have an aperiodic structure. The ultimate path lengths of these wandering level sets are expected to grow further as $s$ increases. The traversal of the level set explored over ${\cal L}=468$ in its three-dimensional projection onto $\phi_1$, $\phi_2$, and $\phi_3$ can be seen as a movie in Figure \ref{biglset} in the online version of this article, with the full structure of the level set shown in the static version of Figure \ref{biglset}. Although numerical searches cannot prove the non-existence of any class of level sets, the numerical simulations did not find level sets that appear to be wandering for $F<1$, which reflects the increased freedom inherent in a high-fluence control field. While a full exploration of the effects of high fluence on allowed level set structures is beyond the scope of this work, the present results show that a rich variety of optimal level set structures exist for sufficiently high control field fluence.

When the fluence was reduced to $F=0.1$, only periodic level sets were found on the landscape. Periodic level sets may be classified as `closed' or `open', depending on the nature of the periodicity. Closed level sets show periodic behavior of each phase, i.e. $\phi_m(s^*)=\phi_m(0),\thickspace\forall m$, while open level sets show growth or decay by 2$\pi n_m$ of at least one $\phi_m$ over a period, i.e., $\phi_m(s^*)=\phi_m(0)\pm2\pi n_m$ for some positive integer $n_m$. Importantly, such an integer 2$\pi$ shift at $s^*$ creates the {\it same} field $\e(0,t)=\e(s^*,t)$ through Eq. (\ref{et}); this behavior is the origin of the label `periodic' for these cases. The difference between closed and open periodic level sets can be visualized two ways, as shown in Figure \ref{opcl} for four distinct level sets on the landscape with $N=3$, $M=5$, and $F=0.1$. Figures \ref{opcl}(a) and (b) show one example of a closed and open level set, respectively, where the value of each $\phi_m$ is plotted versus $s$ over more than two periods. The length of a single period is shown for each level set by the dashed vertical lines at $s^*$ and $2s^*$. In the closed level set of Figure \ref{opcl}(a), all $\phi_m$ satisfy $\phi_m(0)=\phi_m(s^*)=\phi_m(2s^*)$. In the open level set of Figure \ref{opcl}(b), the phase $\phi_1$ (blue) decreases by $\phi_1\to \phi_1- 2\pi$ from $s=0$ to $s=s^*$, and as $s$ is further increased, $\phi_1$ decreases again by $-2\pi$ at $s=2s^*$. The remaining phases in this case of Figure \ref{opcl}(b) satisfy $\phi_l(0)=\phi_l(s^*)=\phi_l(2s^*),\ms l\neq 1$. The origin of the `closed' and `open' terminology is evident when examining projections of these level sets onto the three-dimensional subspace shown in Figure \ref{opcl}(c). The closed level set $c_1$ (from Figure \ref{opcl}(a)), along with the additional cases $c_2$ and $c_3$, create closed curves in the projected space, while the open level set $o_1$ from Figure \ref{opcl}(b) (red) does not close on itself due to the phase $\phi_1$ decreasing by 2$\pi$ over a period. However, projecting the level set $o_1$ onto any three phase controls not including the open control $\phi_1$ makes the level set artificially appear to be a closed curve (not shown). As outlined in Section \ref{lset}, the `size' of a periodic one-dimensional level set may be measured by its path length $\cal L$ from $s=0$ to $s=s^*$. In general, open level sets are larger, as seen by comparing the periods in Figures \ref{opcl}(a) and (b) with the increased size of the open level set in control space, as additionally shown by Figure \ref{opcl}(c). This behavior is also evident in Table \ref{lst}, which presents $\cal L$ for the three open level sets $o_1$, $o_2$, and $o_3$ and three closed level sets $c_1$, $c_2$, and $c_3$ observed at $F=0.1$. 

Of practical interest are the features of optimal level sets as the control field fluence is further reduced below $F=0.1$, because fluence resources are inevitably limited in the laboratory. The effects of decreasing the fluence on the features of optimal level sets are examined beginning from the three periodic open level sets $o_1$, $o_2$ and $o_3$ identified on the landscape for $N=3$, $M=5$, and $F=0.1$. To explore level sets at  values of $F<0.1$, we iteratively decrease the fluence by small discrete steps $\Delta F$. At the $i$th iterative step with fluence $F_i$, an optimal control $\Phi^*(F_i)$ is chosen and the fluence is reduced to $F_{i+1}= F_i-\Delta F_i$. This operation retains the phases $\Phi^*(F_i)$ in Eq. (\ref{et}) for the field, but updates the amplitude $A(t)$ to reflect the new fluence $F_{i+1}$. The latter initial control at $F_{i+1}$ with this procedure produced a yield that is no longer at the top of the landscape, so a gradient climb is then performed beginning at this trial control. Upon satisfaction of $\mpif>0.999999$ with the gradient climb finding a new optimal field at $F_{i+1}$, the second-order search algorithm is employed to map out the nature of the corresponding level set. The iterative procedure for reducing $F$ is repeated until an optimal value $\mpif\geq0.999999$ can no longer be reached. This exploration of optimal level sets at discrete values of $F$ can only provide snapshot glimpses of the fluence dependence of optimal level set features. A thorough investigation would require the development of additional search methods that allow the fluence $F$ to continuously vary as a function of $s$. Nevertheless, the features of the optimal level sets observed here are expected to capture basic aspects of the fluence dependence. 

A schematic diagram labeling the level sets observed at each value of $F$ is shown in Figure \ref{pathlength}(a), and the associated path lengths of each level set are plotted in Figure \ref{pathlength}(b). The solid arrows in Figure \ref{pathlength}(a) and corresponding solid lines in Figure \ref{pathlength}(b) indicate a direct correspondence between level sets (e.g., $o_1\to o_1^\prime$), while dashed arrows and lines indicate suspected {\it combining} and {\it fracturing} of existing level sets. For instance, the three open level sets $o_1^\prime$, $o_2^\prime$, and $o_3^\prime$ at $F=0.095$ appear to {\it combine} to form the closed level set $c_4$ at $F=0.092$. Similarly, the level set $c_4^{\prime\prime}$ at $F=0.088$ appears to {\it fracture} into two closed level sets $c_5$ and $c_6$ at $F=0.085$. Figure \ref{pathlength}(b) shows a wide range of $\cal L$ values for the open level sets at $F\geq0.095$, while the $\cal L$ values decrease rapidly at $F<0.088$. This decrease in $\cal L$ of each level set is magnified in the inset of Figure \ref{pathlength}(b), where ${\cal L}=0$ denotes that the level set has shrunk to an {\it isolated} point at the top of the landscape. The corresponding value of $F$ indicates the minimal `critical' fluence necessary to achieve $P_{1\to 3}\geq0.999999$. 

The presence of distinct critical fluence values from $F=0.069$ through $F=0.077$ at different landscape points illustrates how false traps arise on the landscape as constraints become more severe. At $F=0.069$, the landscape point from level set $c_{12}$ produces an optimal value $P_{1\to 3}=0.999999$, while the four isolated points resulting from the shrinkage of the level sets $c_7^\prime$, $c_9$, $c_{10}$, and $c_{11}$ produce suboptimal $P_{1\to 3}$ values. Gradient climbs with $F=0.069$ beginning from each of the four isolated points were found to become trapped at $P_{1\to 3}$ values ranging from 0.987 through 0.998. Thus, while the optimum can be reached beginning from {\it any} of these five points when $F>0.077$, four of these points become landscape traps when $F$ is reduced to $F=0.069$. From this analysis, the curves in Figure \ref{pathlength}(b) continue to lower values of $P_{1\to 3}$ as points (i.e., ${\cal L}=0$) as the fluence is further lowered for each of them. This behavior links up with the observations in Sections \ref{n2} and \ref{sub} that constraint-induced isolated trapping points readily occur on the landscape. These points are drawn down to suboptimal values from the landscape top as the fluence is lowered.

The combining and fracturing of level sets at the top of the landscape considered in Figure \ref{pathlength} can be visualized by their projections onto three of the phase variables. Figure \ref{lsetflu} plots level sets at selected values of $F\leq0.1$ by their projections onto $\phi_1$, $\phi_4$, and $\phi_5$. The initial open level sets $o_1$ (red), $o_2$ (green), and $o_3$ (blue) at $F=0.1$ are shown in Figure \ref{lsetflu}(a), where the endpoints $s=0$ and $s=s^*$ are shown as colored circles. The corresponding open level sets $o_1^\prime$, $o_2^\prime$, and $o_3^\prime$ at $F=0.095$ are shown in Figure \ref{lsetflu}(b). While $o_1^\prime$ has a similar shape to $o_1$, the level set $o_3^\prime$ only follows part of the $o_3$ trajectory, and has a significantly shorter path length (c.f., Figure \ref{pathlength}(b)). In contrast, $o_2^\prime$ grows in size and can be seen to follow another part of $o_3$ in addition to following $o_2$. This behavior is illustrated schematically by the dashed lines from $o_3$ to $o_2^\prime$ in Figure \ref{pathlength}. Figure \ref{lsetflu}(c) shows that the three open level sets appear to {\it combine} at $F=0.092$ to form level set $c_4$ (cyan), with the corresponding contributing regions of $o_1^\prime$, $o_2^\prime$, and $o_3^\prime$ shown as well. Level set $c_4$ changes little as $F$ is reduced to $F=0.088$, but the resulting level set $c_4^{\prime\prime}$ {\it fractures} into $c_5$ and $c_6$ at $F=0.085$, as shown in Figure \ref{lsetflu}(d). The subsequent fracturing and shrinkage of $c_5$ and $c_6$ at $F\leq0.085$ are shown in Figures \ref{lsetflu}(e) and (f), respectively. 

The optimal level sets explored in this section exhibit many interesting features, both at high and low values of the fluence $F$. Operating at high $F=1$ was found to produce both periodic and wandering level sets on the top of the control landscape, which suggests that a rich variety of optimal level set structures may be present on constrained control landscapes. All observed optimal level sets at $F=0.1$ were found to be periodic, although the numerical methods in this work cannot prove the nonexistence of any class of level sets. The level sets at low fluence values of $F\leq0.1$ appear to undergo both {\it combining} and {\it fracturing} processes, as shown in Figures \ref{pathlength} and \ref{lsetflu}. This result suggests that at some higher value of $F$, the wandering level sets may undergo similar processes to yield periodic level sets at lower values of $F$. Although such combining and fracturing processes cannot be visualized directly from just sampling discrete $F$ values, it can be conjectured that a singularity occurs at a `critical' value of $F$, where either one level set splits into two, or multiple level sets combine into one. The same concept of a `critical' fluence applies when an optimal level set shrinks to an isolated point, which was observed for all of the optimal level sets explored at sufficiently low fluence. This behavior shows that deviations from the anticipated topology of optimal level sets \cite{demiralp,mike2} (i.e., the presence of a one-dimensional critical submanifold when $M=2N-1$) can occur when significant constraints come into play. While these level set features were found here for a particular class of constrained control fields, analogous shrinkage of level sets on control landscapes as fluence is reduced has been observed in quantum control experiments for the objective of molecular fragmentation \cite{expls}. Other control resources (e.g., bandwidth) could play a similar role to fluence when they become a factor limiting the objective yield. These findings provide an impetus for further exploration of optimal and suboptimal level sets, as well as false traps in order to fully understand their features when employing different classes of constrained control fields.

\section{Conclusion}\label{con}
This work explored the topology and local features of constrained quantum control landscapes by choosing a simple parameterization of the control field that provides a small number of physically meaningful controls. The numerical results validated analytical predictions about the topology and structure of optimal solutions  \cite{demiralp,mike2,vinny2}, including the importance of employing at least $2N-2$ (well-chosen) independent controls in order to achieve a $\sim50\%$ probability of reaching the top of the landscape. Suboptimal critical regions on the landscape in the form of both isolated trapping points and saddle regions were identified when the constraints were significant. An increasing prevalence of isolated trapping points was observed as the number of controls was reduced and/or the control field fluence was decreased. Exploration of optimal level sets revealed a rich variety of structures producing $\mpif\sim 1$ at sufficiently high fluence, with the connectedness and size of the level sets decreasing as the fluence was reduced.

The issue of whether traps exist on quantum control landscapes has recently been the subject of much research \cite{mepif,schirmer2,tannor2,comment}. While the presence of traps on otherwise unconstrained control landscapes can be analytically shown for unusual classes of Hamiltonians with constant control fields \cite{schirmer2,tannor2}, extensive numerical simulations with reasonable Hamiltonians and care taken to avoid control constraints have not found any evidence of landscape traps \cite{mepif,mew}. This work takes the further step of demonstrating that a trap-free landscape topology can exist even when a nominally small number of physically reasonable controls is employed. While the landscape topology under {\it any} form of constraints has not been assessed analytically, the results here strongly suggest that the trap-free topology extends even when mild constraints are imposed on the control resources. Furthermore, the appearance of traps at generally high yields (\pif$\gtrsim0.95$) under stronger constraints suggests that the trap-free landscape topology only gradually disappears as constraints are added. The observed lower bound of $2N-2$ controls for relatively easy optimization (i.e., at least $\sim 50\%$ of searches reach the landscape top) is also consistent with analytical results \cite{demiralp,mike2,vinny2}. This `$2N-2$ rule' was found to break down, however, when the controls were poorly chosen. Overall, our results suggest that the inevitable constraints on control field resources that arise in a laboratory setting may not preclude successful optimization.

The presence of optimal level sets on the top of the control landscape is of practical importance because the availability of many optimal solutions makes it possible to select amongst them for secondary characteristics (e.g., robustness of \pif\ms yield to field noise \cite{vinny2}). High control field fluence was found to produce rich optimal level set structures (c.f., Figure \ref{biglset}), however, the field fluence cannot be permitted to grow arbitrarily because additional physical processes may enter, including possibly of an undesirable nature. Reducing the fluence led to fracturing of the level sets and their ultimate shrinkage to isolated points that pulled away from the top of the landscape to form false traps as the fluence continued to decrease. The placement of these analyses in the context of non-linear dynamical phenomena \cite{guckenheimer,strogatz} opens up a new direction for assessing control landscape features. Many other measures of the control field can also affect the features of optimal and suboptimal solutions.

This work presented a systematic study of the effects of control constraints on the quantum control landscape for pure-state population transfer. The gradual retreat from the ideal trap-free topology observed as the control resources become more limited helps to explain the general success of many quantum control experiments even when employing constrained controls with limited laser bandwidth and pulse energy \cite{Weinacht2005b,melab2}. In particular, analogous optimal level set features to those found in this paper were observed in molecular fragmentation experiments as the control field fluence was reduced \cite{expls}. In addition to the phase controls employed here, other control field parameterizations need to be explored in both simulations and experiments with the aim of identifying physically reasonable control bases that optimize a broad variety of quantum control objectives.

\acknowledgements{The authors acknowledge support from the NSF, ARO, and DOE. K.W.M. acknowledges the support of an NSF graduate research fellowship.}

\bibliography{constraint}

\begin{table}[htbp]
\caption{Path lengths $\cal L$ of optimal level sets for $N=3$, $M=5$, with fluence $F=1$ and $F=0.1$. The labels for the $F=0.1$ level sets are presented in the text, as well as in the Figure(s) indicated. The designations `open' and `closed' denote the type of periodicity and are explained Sections \ref{lset} and \ref{lsetexp}. The `wandering' level set at $F=1$ does not appear to have a periodic structure, as explained Section \ref{lsetexp}.\label{lst}}
 \begin{tabular}{rcccr}
\hline\hline
  $F$&label&Figure&type&$\cal L$\\\hline
1&&\ref{biglset}&wandering&$>468$\\
&&&wandering&$>137$\\
&&&wandering&$>93$\\
&&&closed&1.2\\
&&&closed&7.8\\
0.1&$c_1$&\ref{opcl}(a) and (c)&closed&4.5\\
&$c_2$&\ref{opcl}(c)&closed&8.1\\
&$c_3$&\ref{opcl}(c)&closed&7.2\\
&$o_1$&\ref{opcl}(b), (c), and \ref{lsetflu}(a)&open&21.6\\
&$o_2$&\ref{lsetflu}(a)&open&25.1\\
&$o_3$&\ref{lsetflu}(a)&open&135\\\hline\hline
 \end{tabular}
\end{table}

\section*{Figure captions}

\begin{figure}[htbp]
\includegraphics[width=8.5cm]{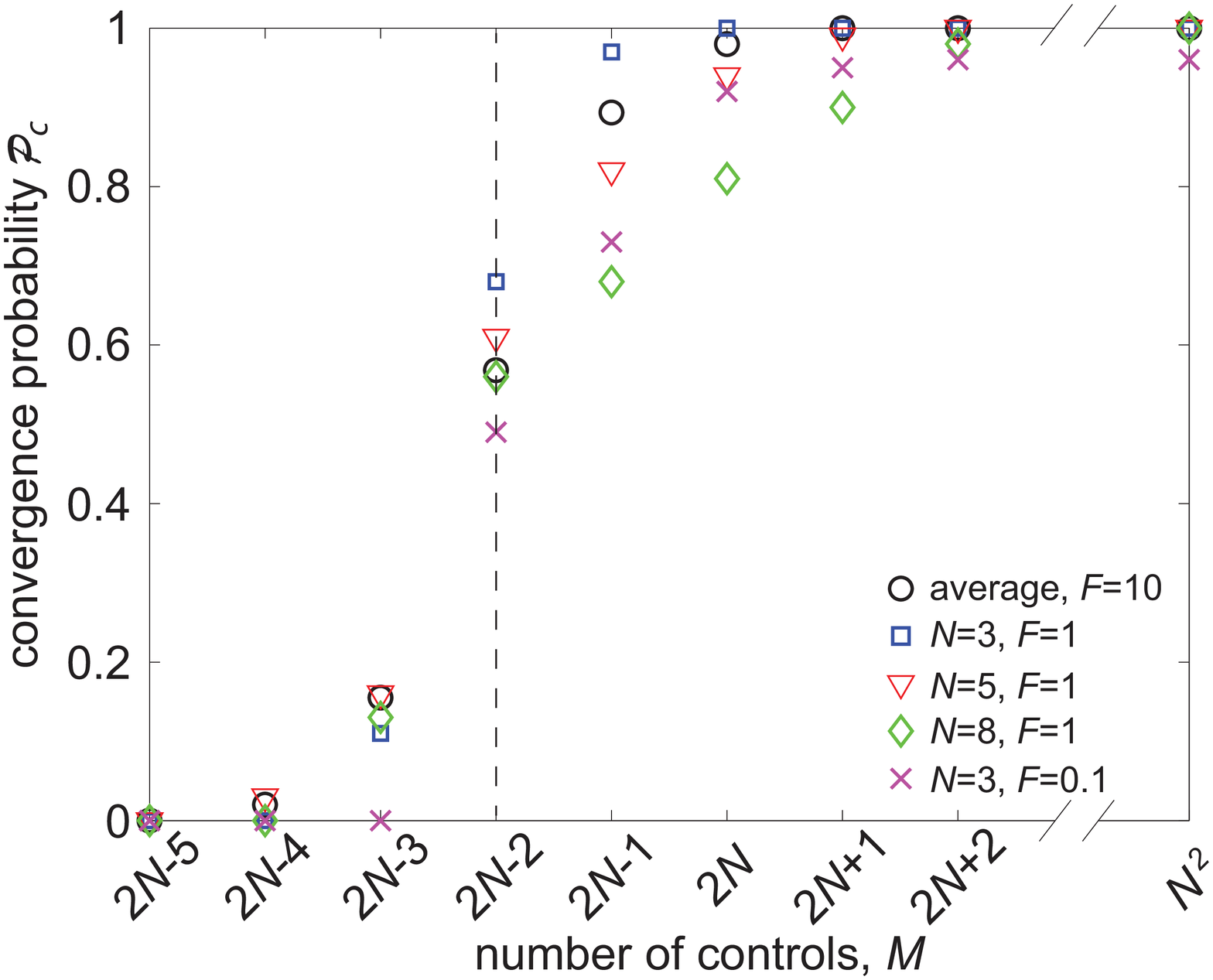}
\caption{ \label{convstat}}
\end{figure}

Figure \ref{convstat}: Convergence probability ${\cal P}_c$ of optimization searches reaching the threshold $P_{1\to N}>0.999$ versus the number of controls $M$. The average value of ${\cal P}_c$ over $N$ from 3 through 8 is shown for $F=10$ (black circles) because ${\cal P}_c$ was found to be essentially the same for all $N$ from 3 to 8. The ${\cal P}_c$ value is shown for $N=3$ (blue squares), $N=5$ (red triangles), and $N=8$ (green diamonds) at $F=1$, with ${\cal P}_c$ decreasing as $N$ rises. ${\cal P}_c$ is shown for $F=0.1$, $N=3$ (magenta x), and not all searches converge even when $M=N^2$. The dashed vertical line denotes $M=2N-2$ controls.

\begin{figure}[htbp]
\includegraphics[width=8.5cm]{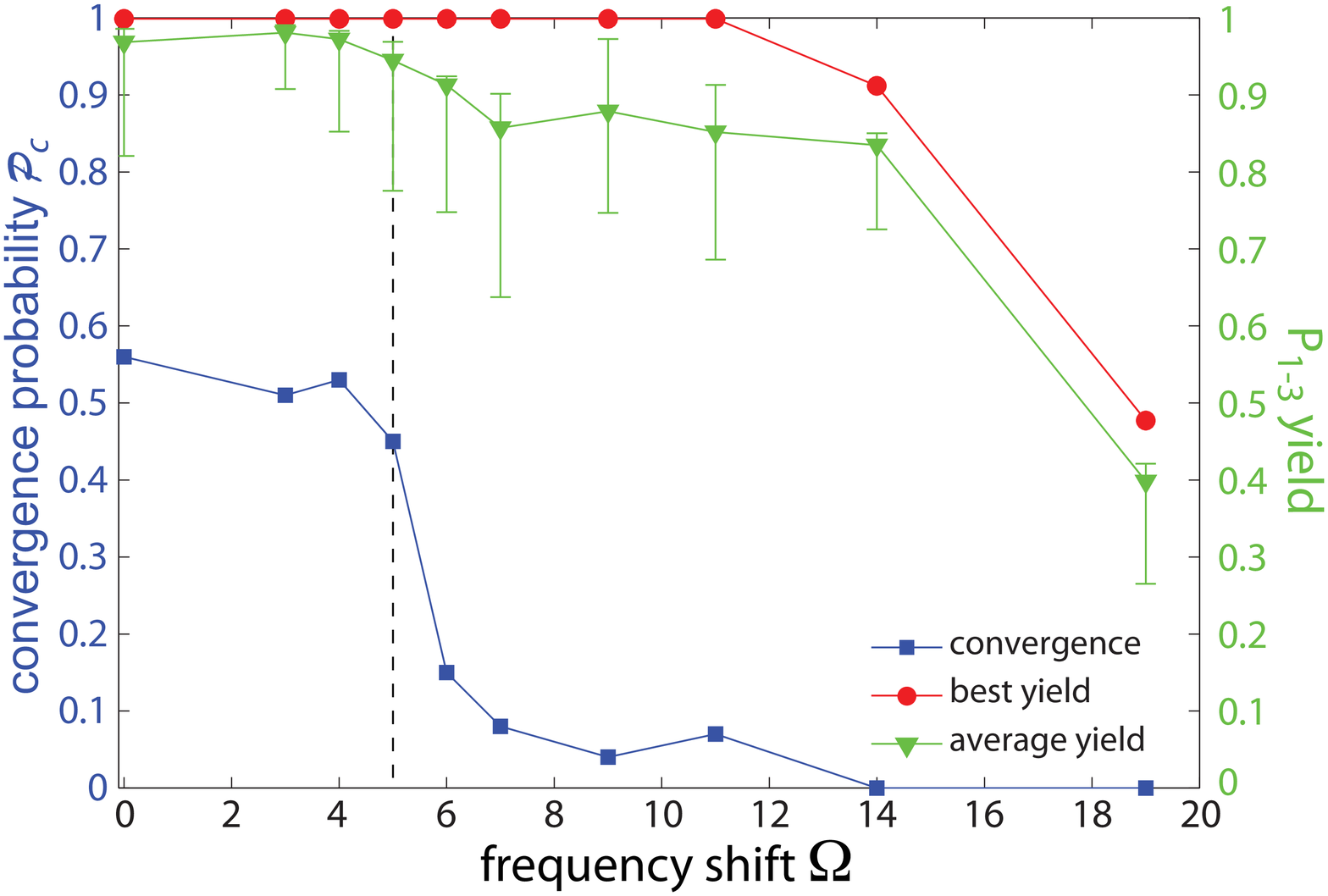}
\caption{\label{omega}}
\end{figure}

Figure \ref{omega}: Convergence statistics for $N=3$, $M=4$, $F=10$, with the control field frequencies in Eq. (\ref{et}) shifted to $\omega_m=m+\Omega$, plotted as a function of $\Omega$: probability ${\cal P}_c$ of reaching $P_{1\to3}>0.999$ (blue squares), maximal $P_{1\to 3}$ yield (red circles), and mean yield $\langle P_{1\to 3}\rangle$ with error bars denoting left and right standard deviation (green triangles). The values of ${\cal P}_c$ and $\langle P_{1\to 3}\rangle$ decrease as $\Omega$ grows beyond $\Omega+1>H_0(3,3)-H_0(1,1)=6$, denoted by the vertical dashed line at $\Omega=5$.

\begin{figure}[htbp]
\includegraphics[width=8.5cm]{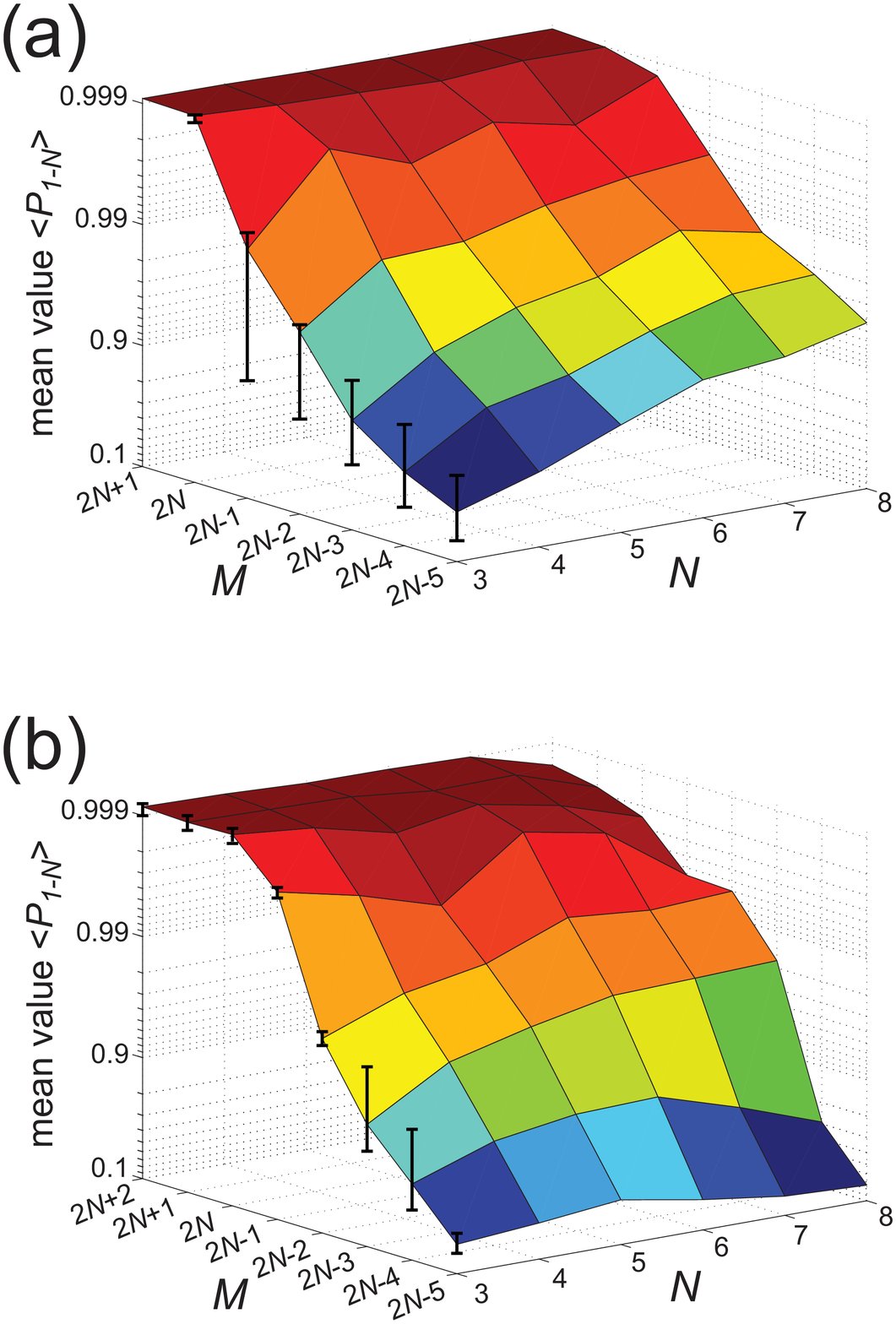}
\caption{ \label{d3}}
\end{figure}

Figure \ref{d3}: Mean value $\langle P_{1\to N}\rangle$ from 100 runs versus the number of controls $M$ and system dimension $N$ for $F=10$ (a) and $F=1$ (b). Error bars denoting the left and right standard deviations from the mean are included for $N=3$. In (a), the value $\langle P_{1\to N}\rangle$ rises both with the number of controls $M$ and $N$, while in (b), $\langle P_{1\to N}\rangle$ decreases for some cases of $M$ as $N$ grows. This behavior is consistent with limited fluence imposing a further constraint in (b) but not in (a) with its higher fluence value.

\begin{figure}[htbp]
 \includegraphics[width=8.5cm]{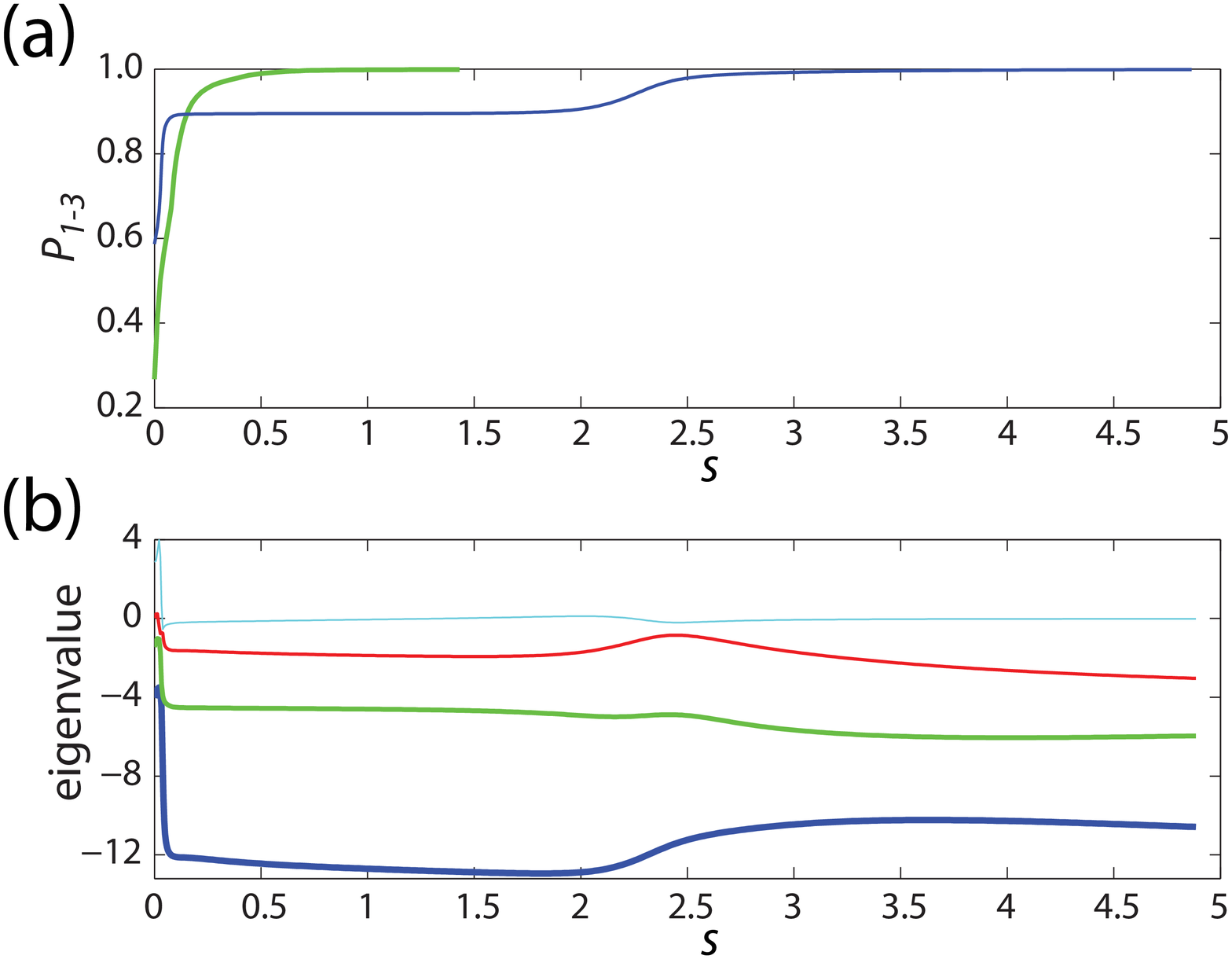}
\caption{\label{saddlef}}
\end{figure}

Figure \ref{saddlef}: Search trajectory passing through a saddle point, compared to a trajectory that does not encounter a saddle point, for $N=3$, $M=4$, and $F=1$. The $P_{1\to 3}(s)$ value for the search trajectory encountering a saddle (blue) and the trajectory not encountering a saddle (green) are shown as a function of $s$ in (a). For the trajectory that encounters a saddle, the four Hessian eigenvalues as a function of $s$ are shown in (b). In the saddle region, the eigenvalue denoted by the cyan line has a very small positive value of 0.003 at $s=0.91$, where the $P_{1\to 3}$ trajectory is flattest.

\begin{figure}[htbp]
\includegraphics[width=8.5cm]{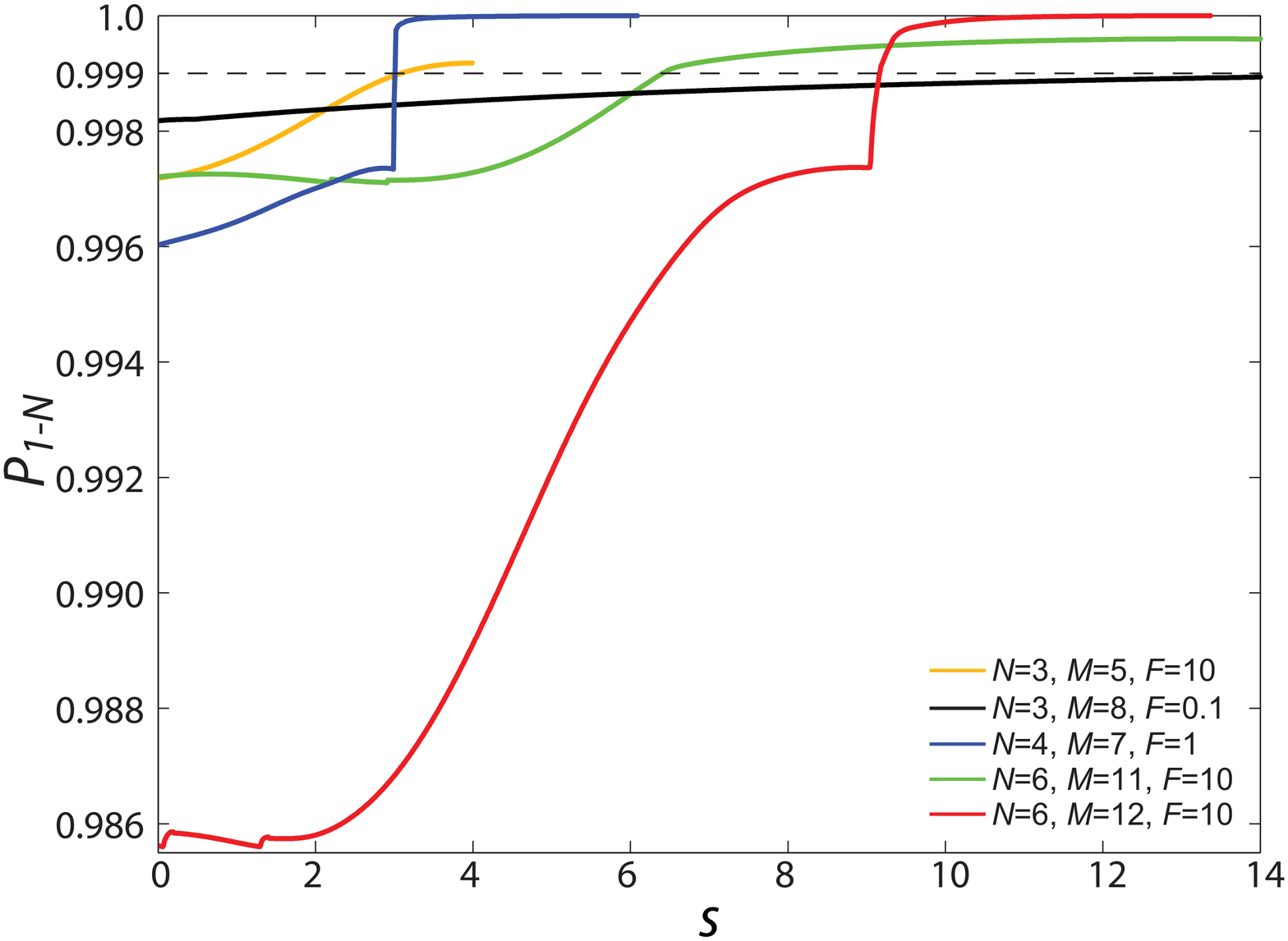}
\caption{ \label{trapsaddle}}
\end{figure}

Figure \ref{trapsaddle}: Search trajectories of $P_{1\to N}$ versus index $s$ for apparent constrained landscape traps that are in fact saddles. The $P_{1\to N}$ trajectories as a function of $s$ escape from the initial putative trapped $P_{1\to N}^*$ values at $s=0$, and the optimal value $P_{1\to N}\geq0.999$ (shown by the dashed horizontal line) is attained. The sudden jumps in \pif\ms value on the blue and red trajectories occur due to switching from following the Hessian eigenvector corresponding to the smallest magnitude eigenvalue to following the gradient. These searches converged to $P_{1\to N}=0.999999$. On the green and red curves, the lowest $P_{1\to N}$ value was the threshold $P_{1\to N}^\star-0.0001$ (see the discussion in the text).

\begin{figure}[htbp]
 \includegraphics[width=8.5cm]{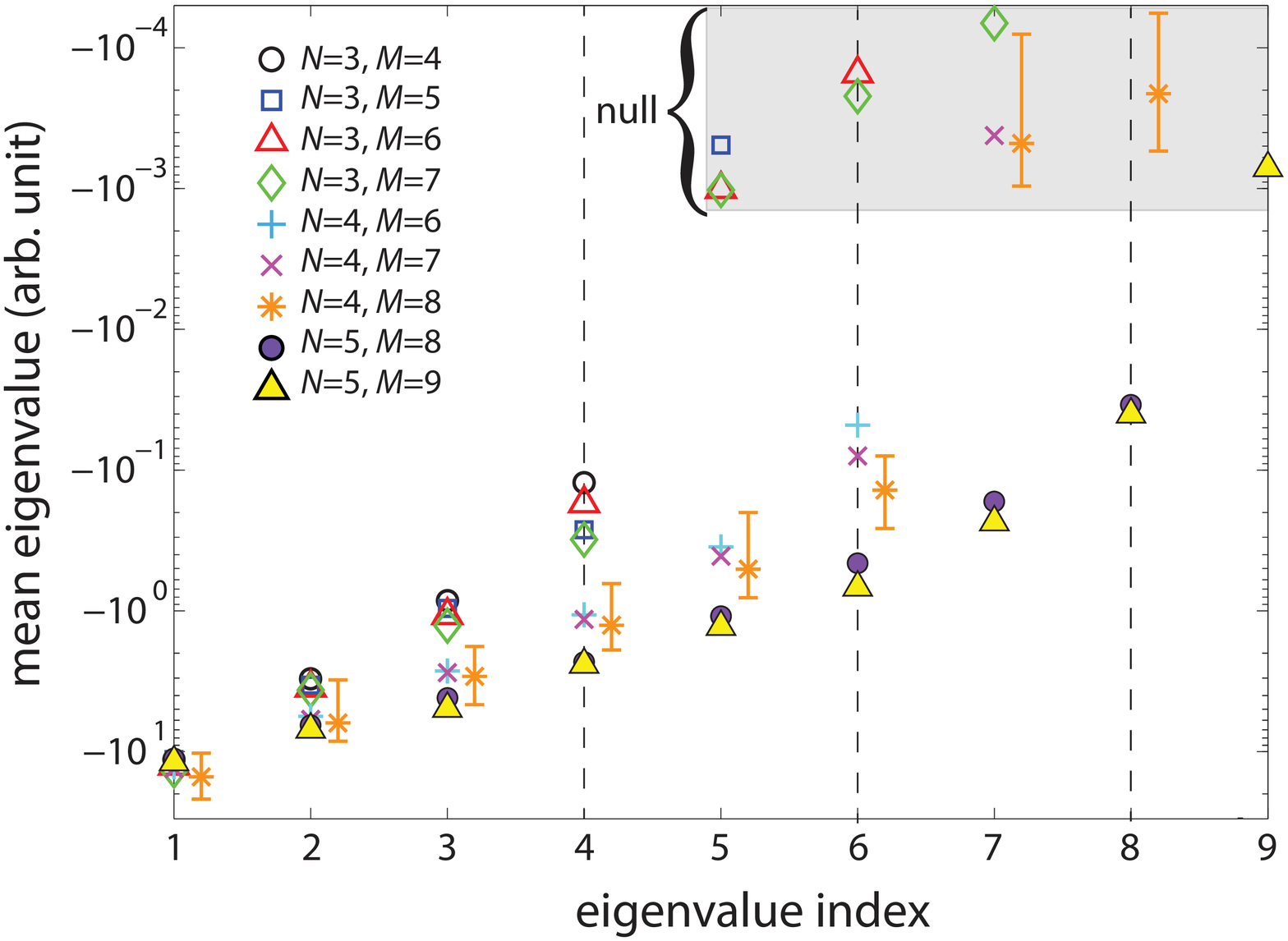}
\caption{ \label{eigs}}
\end{figure}

Figure \ref{eigs}: Mean Hessian eigenvalues versus their index at the top of the landscape where $\mpif\geq0.999999$. The index of $2N-2$ is shown for $N=3$, 4, and 5 by the respective dashed lines. Eigenvalues below their associated $2N-2$ index are distinctly negative, on average $\lesssim-0.1$. Eigenvalues above this index are found to be $\gtrsim-0.002$ and considered as null; these are labeled by the domain with a brace and highlighted in grey on the figure. Error bars denoting the left and right standard deviations from the mean are shown for the representative example of $N=4$, $M=8$, where the points are shifted on the abscissa for graphical clarity. Error bars for the other cases were of similar magnitude.

\begin{figure}[htbp]
\includegraphics[width=8.5cm]{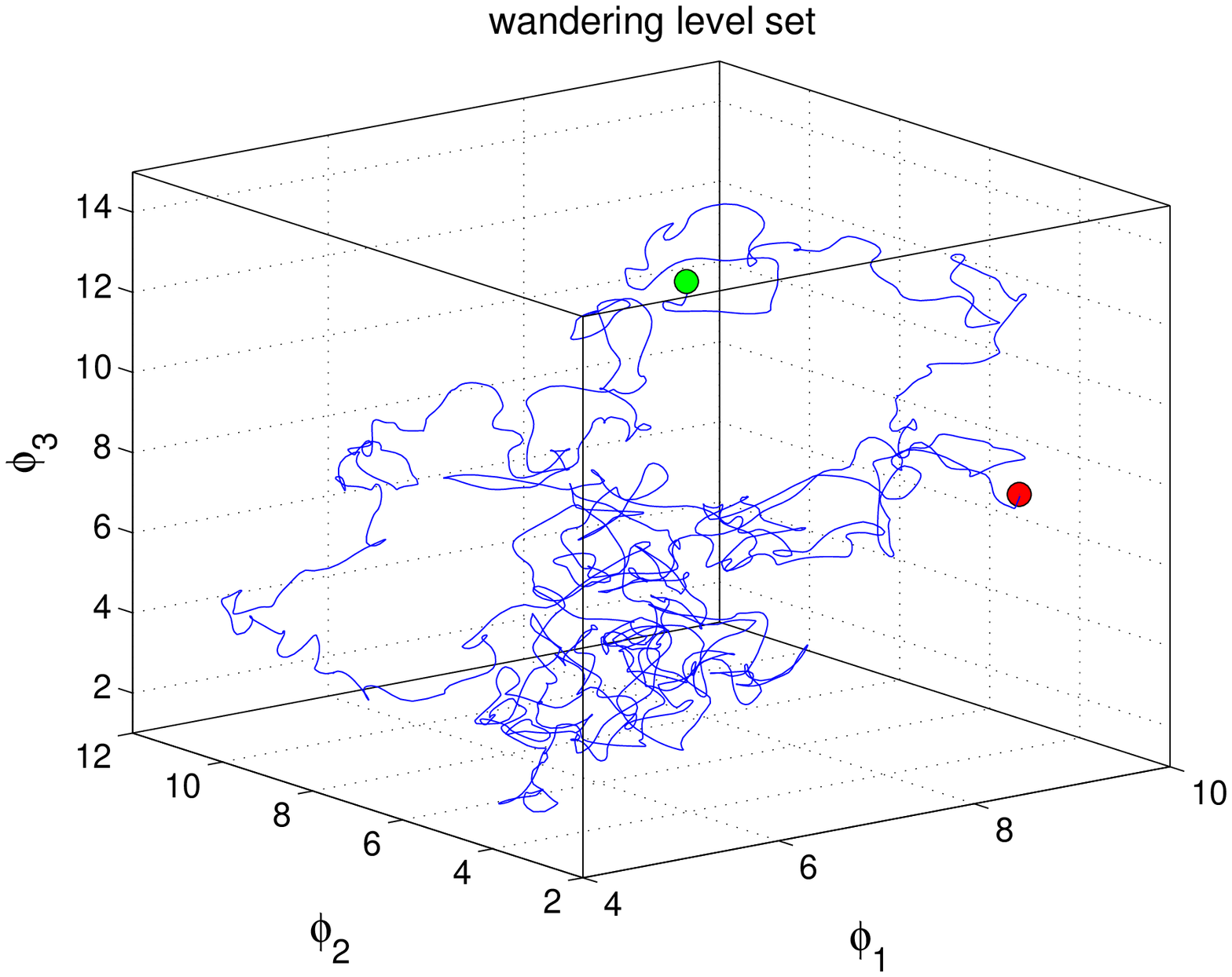}
\caption{ \label{biglset}}
\end{figure}

Figure \ref{biglset}: (enhanced online) Projection of a wandering level set at the top of the landscape for $N=3$, $M=5$, and $F=1$ onto $\phi_1$, $\phi_2$, and $\phi_3$. The green and red dots on the still image denote the beginning and end, respectively, of the trajectory explored by the second-order search procedure of Section \ref{lset}. This level set was followed over more than 10$^6$ $s$-steps, with a corresponding path length of ${\cal L}=468$, and the level set is expected to extend beyond the region explored here as $s$ increases further. This level set is called `wandering' in Table \ref{lst}, as the same physical point in search space (i.e., producing an identical $\e(t)$) was not encountered twice over the trajectory of $M=5$ variables. The traversal of this level set by the red dot is animated in the associated movie.

\begin{figure}[htbp]
 \includegraphics[width=8.5cm]{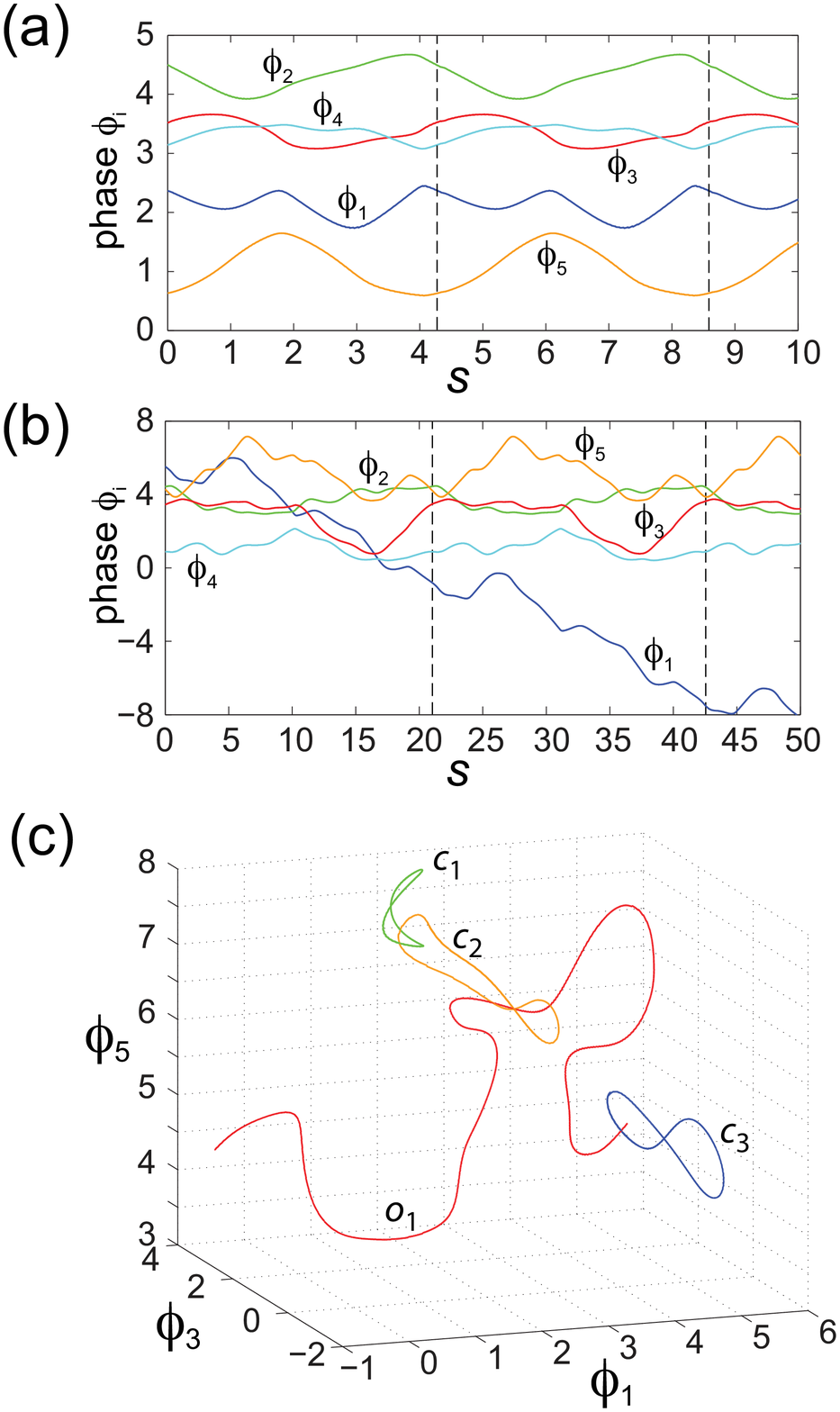}
\caption{\label{opcl}}
\end{figure}

Figure \ref{opcl}: Phase variables plotted versus $s$ at the top of the landscape for (a) a periodic closed level set $c_1$ and (b) a periodic open level set $o_1$. These are two disconnected level sets on the landscape for $N=3$, $M=5$, and $F=0.1$. The dashed vertical lines denote the values $s^*$ and $2s^*$, where the first period of each level set is enclosed in $0\leq s\leq s^*$, and the second period is over $s^*\leq s\leq 2s^*$. In the closed level set, all phases return to their initial values at $s=s^*$ and $s=2s^*$. In the open level set case shown here, the phase $\phi_1$ (blue) decreases by a factor of 2$\pi$ over one period, while the other phases return to their initial values at $s=0$. In (c), a three-dimensional projections onto $\phi_1$, $\phi_3$, and $\phi_5$ of three closed level sets. Here, $c_1$ is the closed level set in (a), while $c_2$, and $c_3$ are other closed level sets. The open level set $o_1$ is that in (b). The closed level sets form closed curves in search space, while the open level set does not close on itself. However, if the open level set $o_1$ is projected onto three of the phase controls that do $not$ include the `open' control $\phi_1$, then this level set artificially appears as a closed curve as well.

\begin{figure}[htbp]
\includegraphics[width=8.5cm]{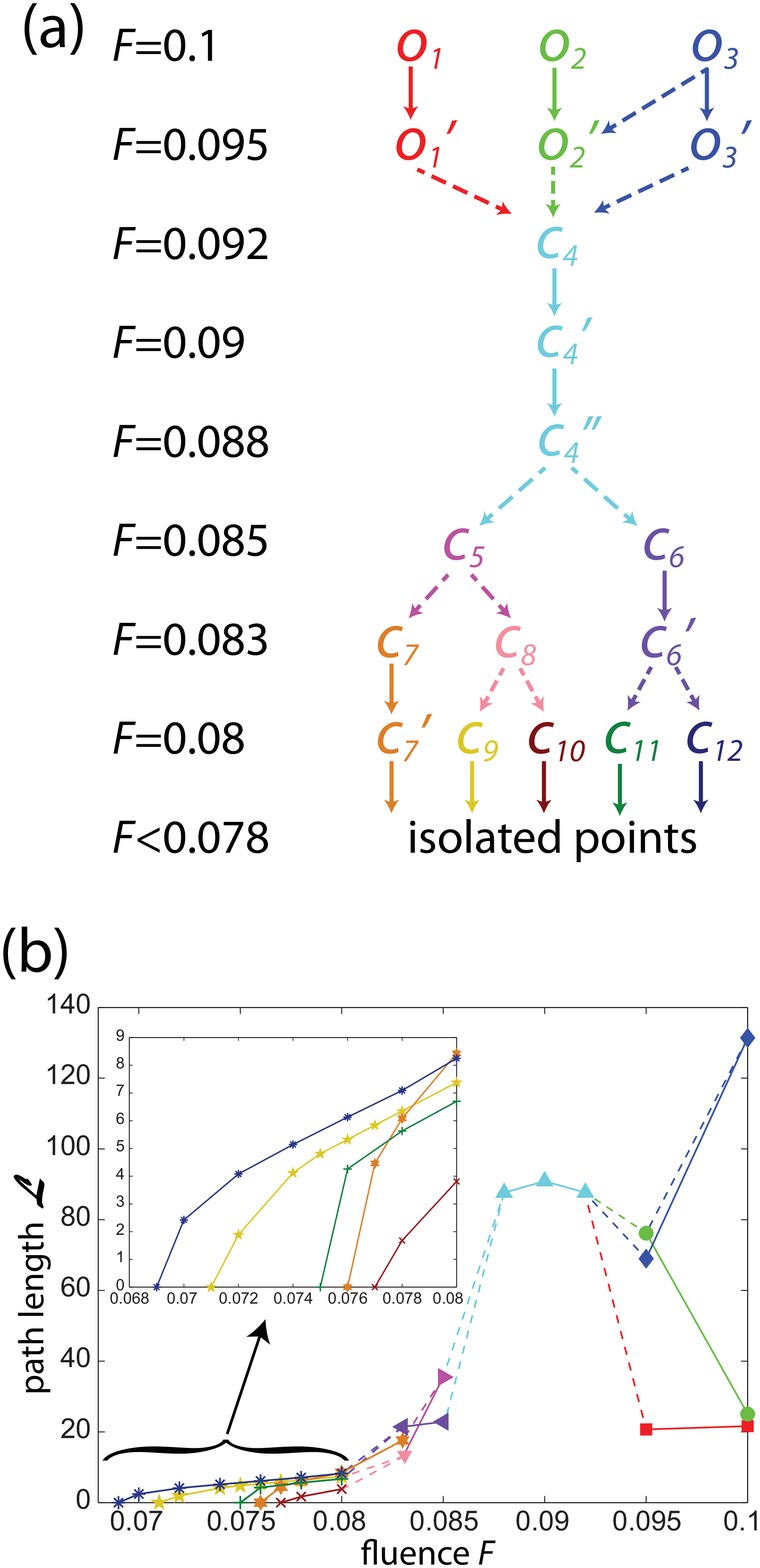}
\caption{\label{pathlength} }
\end{figure}

\begin{figure}[htbp]
\includegraphics[width=15cm]{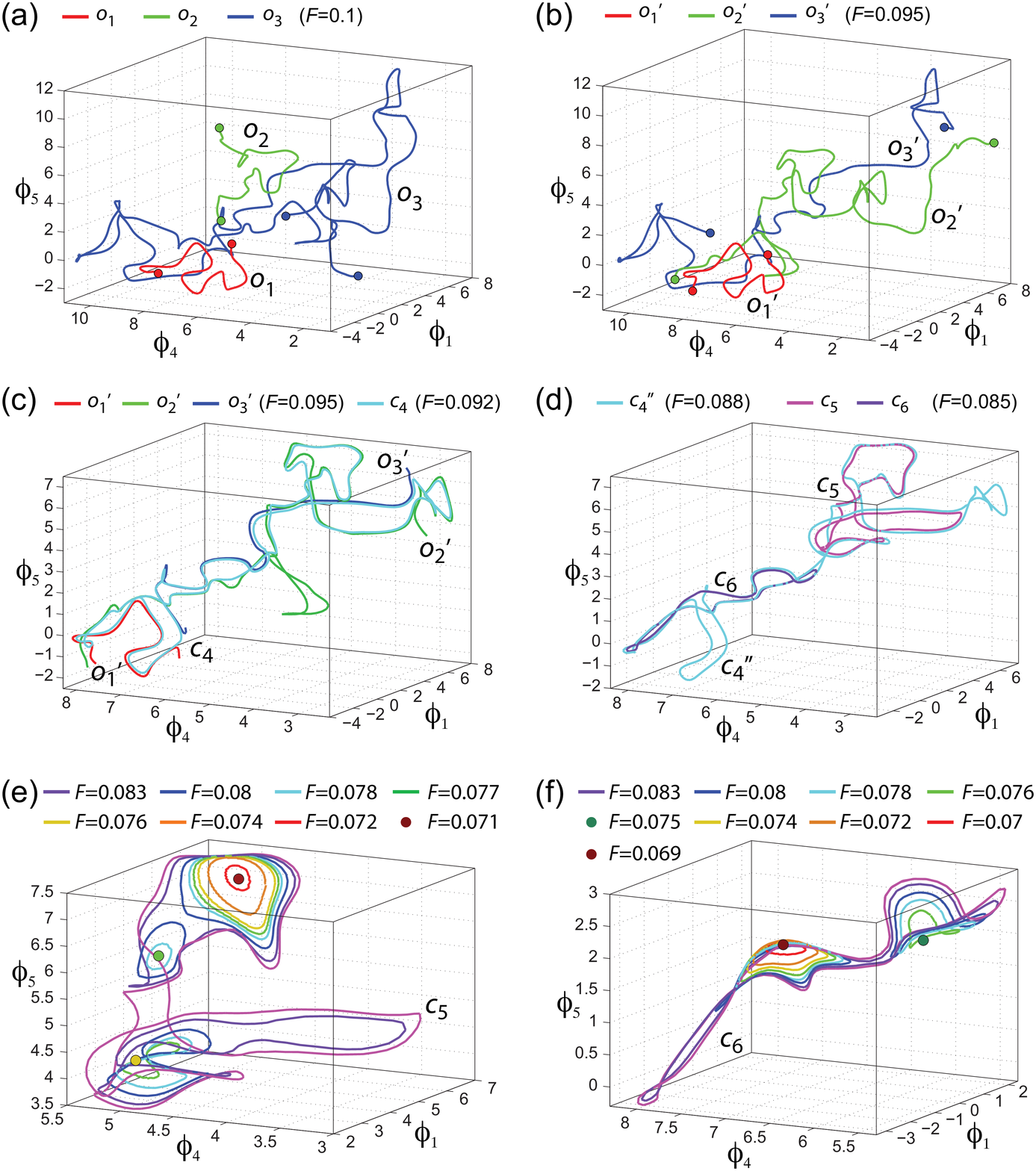}
\caption{\label{lsetflu}}
\end{figure}

Figure \ref{pathlength}: Illustration of optimal level set features as a function of control field fluence $F$ (see the text for details). (a) Schematic diagram of level sets observed at each value of $F$. (b) Plot of the path length $\cal L$ for each level set as a function of the fluence $F$. The solid lines denote direct correspondence between level sets as $F$ decreases. The dashed lines denote suspected combining or fracturing of the level sets. The inset shows the shrinkage of the five level sets $c_7^\prime$, $c_9$, $c_{10}$, $c_{11}$, and $c_{12}$ at $F\leq0.08$. These five level sets shrink to isolated points with ${\cal L}=0$ at distinct values of $F$ ranging from $F=0.069$ through $F=0.077$. Further reduction of $F$ in each case takes these isolated points off the top of the landscape to become suboptimal constraint-induced traps. The diagram in (a) is also shown in the animated movie of Figure \ref{lsetflu} in the online version of this article.

Figure \ref{lsetflu}: (enhanced online) Projections of optimal level sets for $N=3$ and $M=5$ onto the phase controls $\phi_1$, $\phi_4$, and $\phi_5$ at selected values of the fluence $F\leq0.1$. The level sets are animated in the associated movie as $F$ decreases; the six plots (a) through (f) shown in the still image highlight some important features: (a) The three open level sets $o_1$ (red; this trajectory is distinct from Figure \ref{opcl}(c), as $\phi_4$ is shown), $o_2$ (green), and $o_3$ (blue) at $F=0.1$. (b) The corresponding level sets $o_1^\prime$, $o_2^\prime$, and $o_3^\prime$ at $F=0.095$. (c) The closed level set $c_4$ at $F=0.092$ (cyan), along with the corresponding regions of the three level sets at $F=0.095$ that combine to form $c_4$. (d) Level set $c_4^{\prime\prime}$ at $F=0.088$ (cyan), and the level sets $c_5$ (magenta) and $c_6$ (violet) into which it fractures at $F=0.085$. (e) Level set $c_5$ and its subsequent fracturing at $F<0.085$. (f) Level set $c_6$ and its subsequent fracturing at $F<0.085$. Further reduction of $F$ shrinks all of the level sets to isolated points at different values of $F$ as shown by the colored circles in (e) and (f), which are highlighted in red in the associated animation. Upon even further reduction of $F$, these isolated points on the landscape `fall off' the landscape top to become isolated constraint-induced trapping points.

\newpage

\end{document}